\documentclass{cernrep}
\usepackage{graphicx,here}
 
\begin{document}

\title{A QCD PRIMER}
 
\author{G. Altarelli}
 
\institute{CERN, Geneva, Switzerland}
%-----------------------------------------------------------------------
% If your printer does not reproduce dimensions exactly, it may be
% necessary to remove the % signs and adjust the dimensions in the
% following commands:
%
%     \setlength{\textheight}{24cm}
%     \setlength{\textwidth}{16cm}
%
% Similarly for the following, if you need to adjust the positioning
% on the paper:
%
%         \setlength{\topmargin}{-1cm}
%         \setlength{\oddsidemargin}{0pt}
%         \setlength{\evensidemargin}{0pt}
%------------------------------------------------------------------------
%DEFINITIONS OF THE AUTHOR
\def\beq{\begin{equation}} 
\def\eeq{\end{equation}} 
\def\bea{\begin{eqnarray}} 
\def\eea{\end{eqnarray}}
\def\bq{\begin{quote}} 
\def\eq{\end{quote}}

\def\AJ{{\it Astrophys.J.} } 
\def\AJL{{\it Ap.J.Lett.} } 
\def\AJS{{\it Ap.J.Supp.} } 
\def\AM{{\it Ann.Math.} } 
\def\AP{{\it Ann.Phys.} } 
\def\APJ{{\it Ap.J.} } 
\def\APP{{\it Acta Phys.Pol.} }
\def\ASAS{{\it Astron. and Astrophys.} } 
\def\BAMS{{\it Bull.Am.Math.Soc.} } 
\def\CMJ{{\it Czech.Math.J.} } 
\def\CMP{{\it Commun.Math.Phys.} } 
\def\FP{{\it Fortschr.Physik} } 
\def\HPA{{\it Helv.Phys.Acta} } 
\def\IJMP{{\it Int.J.Mod.Phys.} } 
\def\JMM{{\it J.Math.Mech.} } 
\def\JP{{\it J.Phys.} } 
\def\JCP{{\it J.Chem.Phys.} } 
\def\LNC{{\it Lett. Nuovo Cimento} } 
\def\SNC{{\it Suppl. Nuovo Cimento} } 
\def\MPL{{\it Mod.Phys.Lett.} } 
\def\NAT{{\it Nature} } 
\def\NC{{\it Nuovo Cimento} }
\def\NP{{\it Nucl.Phys.} } 
\def\PL{{\it Phys.Lett.} } 
\def\PR{{\it Phys.Rev.} } 
\def\PRL{{\it Phys.Rev.Lett.} } 
\def\PRTS{{\it Physics Reports} } 
\def\PS{{\it Physica Scripta} } 
\def\PTP{{\it Progr.Theor.Phys.} } 
\def\RMPA{{\it Rev.Math.Pure Appl.} } 
\def\RNC{{\it Rivista del Nuovo Cimento} }
\def\SJPN{{\it Soviet J.Part.Nucl.} } 
\def\SP{{\it Soviet.Phys.} } 
\def\TMF{{\it Teor.Mat.Fiz.} }
\def\TMP{{\it Theor.Math.Phys.} } 
\def\YF{{\it Yadernaya Fizika} } 
\def\ZETF{{\it Zh.Eksp.Teor.Fiz.} }
\def\ZP{{\it Z.Phys.} } 
\def\ZMP{{\it Z.Math.Phys.} }

\def\gappeq{\mathrel{\rlap {\raise.5ex\hbox{$>$}} {\lower.5ex\hbox{$\sim$}}}}

\def\lappeq{\mathrel{\rlap{\raise.5ex\hbox{$<$}} {\lower.5ex\hbox{$\sim$}}}}
%%%%%%%%%%%%%%%%%%%%%%%%%%%%%%%%%%%%%%%%%%%%%%%%%%%%%%%%%%%%%%%%%%%%%%%%%%%%%%% 

\maketitle % this produces the title block
 
\begin{abstract}
1. Introduction\\ 
2. Massless QCD and Scale Invariance \\ 
3. The Renormalisation Group and Asymptotic Freedom\\ 
4. More on the Running Coupling\\ 
5. Application to Hard Processes\\   
\hphantom{5.~~}5.1 $R_{e^+e^-}$ and Related Processes\\  
\hphantom{5.~~}5.2 The Final State in $e^+e^-$ Annihilation\\ 
\hphantom{5.~~}5.3 Deep Inelastic Scattering\\  
\hphantom{5.~~}5.4 Factorisation and the QCD Improved Parton Model\\ 
6. Measurements of $\alpha_s$\\   
\hphantom{6.~~}6.1 $\alpha_s$ from $e^+e^-$ Colliders\\   
\hphantom{6.~~}6.2 $\alpha_s$ from Deep Inelastic Scattering\\
7. Conclusion\\ 
8. Appendix: The Formalism of Gauge Theories 
\end{abstract}
 
\section{INTRODUCTION}
These four lectures are devoted to an elementary introduction to Quantum Chromo-Dynamics (QCD), the theory of strong
interactions. Four lectures are not much. So after a general introduction I will concentrate on the basic principles and
the main applications of perturbative QCD (for reviews of the subject see for example \cite{PR}, \cite{AR}, \cite{Book}). I
will try to put the main emphasis on ideas with only a minimum of technicalities.

At present most of particle physics is well understood in terms of the Standard Model (SM), which is a gauge theory
of strong and electroweak interactions based on the group $SU(3)\bigotimes SU(2)\bigotimes U(1)_Y$. The $SU(3)$ factor is
the colour group of QCD, while $SU(2)\bigotimes U(1)_Y$ is the Glashow-Weinberg-Salam electroweak symmetry group. The
electroweak symmetry is spontaneously broken down to $U(1)_Q$, the phase group of the electric charge $Q_e$ (that is, $Q$ is the charge operator in units of $e$, the proton charge), different from
the $U(1)_Y$ of weak hypercharge: $Q=t_3+Y/2$, where $t_3$ is the third component of the weak isospin generator of
$SU(2)$. The group $SU(3)\bigotimes U(1)_Q$ is believed to be an exact gauge symmetry of nature. The corresponding gauge
bosons, the eight gluons and the photon are massless. Matter fields include three families of coloured quarks and
colourless leptons. Quarks and gluons (q and g) are the fields that have strong interactions described by QCD. The
statement that QCD is a gauge theory based on the group $SU(3)$ with colour triplet quark matter fields fixes the QCD
lagrangian density to be (a summary of the general formalism of gauge theories is presented in Appendix):
\beq
{\cal L}~=~-\frac{1}{4}\sum_{A=1}^8F^{A\mu\nu}F^A_{\mu\nu}~+~\sum_{j=1}^{n_f}\bar q_j(iD\llap{$/$}-m_j)q_j\label{LagQCD}\\
\eeq
Here: $q_j$ are the quark fields (of $n_f$ different flavours) with mass $m_j$; $D\llap{$/$}=D_{\mu}\gamma^{\mu}$, where
$\gamma^{\mu}$ are the Dirac matrices and $D_{\mu}$ is the covariant derivative: 
\beq
D_{\mu}=\partial_{\mu}-ie_s{\bf g_{\mu}};\label{der}\\
\eeq $e_s$ is the gauge coupling, later we will mostly use, in analogy with QED
\beq
\alpha_s=\frac{e_s^2}{4\pi};\label{alfa}\\
\eeq
${\bf g_{\mu}}= \sum_A~t^Ag_{\mu}^A~$
where
$g_{\mu}^A$, $A=1,8$, are the gluon fields and
$t^A$ are the
$SU(3)$ group generators in the triplet representation of quarks (i.e. $t_A$ are 3x3 matrices acting on $q$); the
generators obey the commutation relations  $[t^A,t^B]=iC_{ABC}t^C$ where $C_{ABC}$ are the complete antisymmetric
structure constants of $SU(3)$ (the normalisation of $C_{ABC}$ and of $e_s$ is specified by
$Tr[t^At^B]=\delta^{AB}/2$);
\beq
F^A_{\mu\nu}~=~\partial_{\mu} g^A_{\nu}-\partial_{\nu} g^A_{\mu}~-~e_sC_{ABC}g^B_{\mu}g^C_{\nu}\label{F}
\eeq

The Feynman rules of QCD are listed in Fig~1. The physical vertices in QCD include the gluon-quark-antiquark vertex,
analogous to the QED photon-fermion-antifermion coupling, but also the 3-gluon and 4-gluon vertices, of order $e_s$ and
$e_s^2$ respectively, which have no analogue in an abelian theory like QED. In fact the QED lagrangian density is given
by:

\begin{figure}
\includegraphics[width=0.95\textwidth,clip]{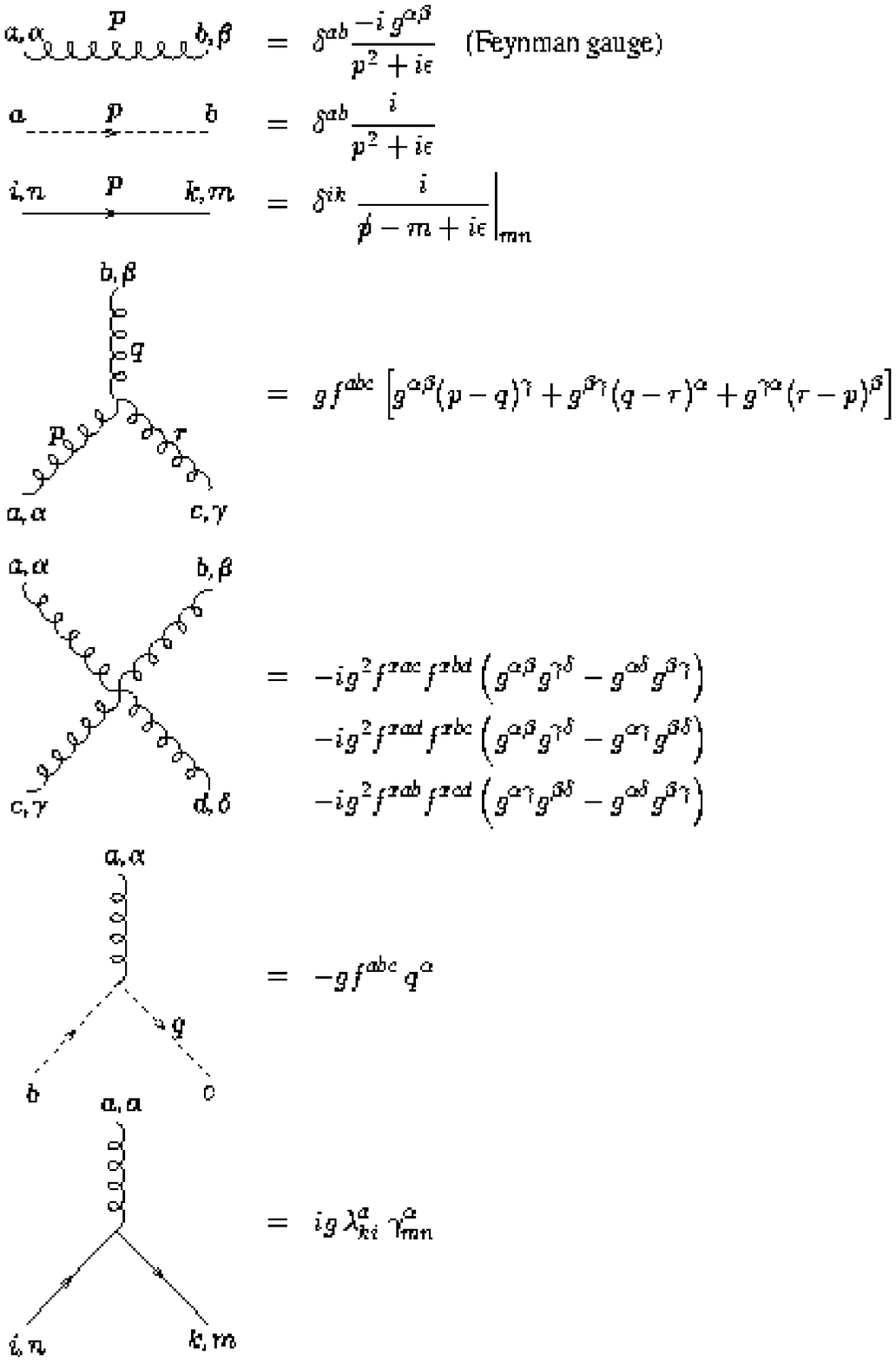} 
\caption{Feynman rules for QCD. The solid lines
 represent the femions, the curly lines the gluons, and the dotted
 lines represent the ghosts.}
\end{figure}
 
\beq
{\cal L}~=~-\frac{1}{4}F^{\mu\nu}F_{\mu\nu}~+~\sum_{\psi}\bar {\psi}(iD\llap{$/$}-m_{\psi})\psi\label{LQED}\\
\eeq  
with the covariant derivative given in terms of the photon field $A_{\mu}$ and the charge operator Q by: 
\beq
D_{\mu}=\partial_{\mu}-ieA_{\mu}Q\label{derQED}\\
\eeq
and 
\beq
F_{\mu\nu}~=~\partial_{\mu} A_{\nu}-\partial_{\nu} A_{\mu}\label{FQED}\\
\eeq
In QED the photon is coupled to all electrically charged particles but itself is neutral. In QCD the gluons are coloured
hence self-coupled. This is reflected in the fact that in QED $F_{\mu\nu}$ is linear in the gauge field, so that the term
$F_{\mu\nu}^2$ in the lagrangian is a pure kinetic term, while in QCD $F^A_{\mu\nu}$ is quadratic in the gauge field so
that in
$F^{A2}_{\mu\nu}$ we find cubic and quartic vertices beyond the kinetic term. Also instructive is to consider the case of
scalar QED: 
\beq
{\cal L}~=~-\frac{1}{4}F^{\mu\nu}F_{\mu\nu}~+~(D_{\mu}\phi)^\dagger (D^{\mu}\phi)-m^2(\phi^\dagger \phi)\label{LSQED}\\
\eeq
For $Q=1$ we have:
\beq
(D_{\mu}\phi)^\dagger (D^{\mu}\phi)~=~(\partial_{\mu}\phi)^\dagger (\partial^{\mu}\phi)~+~ieA_{\mu}[(\partial^{\mu}\phi)^\dagger \phi
~-~\phi^\dagger (\partial^{\mu}\phi)]~+~e^2A_{\mu}A^{\mu}\phi^\dagger \phi
\label{seagull}\\
\eeq
We see that for a charged boson in QED, given that the kinetic term for bosons is quadratic in the derivative, there is a
two-gauge vertex of order
$e^2$. Thus in QCD the 3-gluon vertex is there because the gluon is coloured and the 4-gluon vertex because the gluon is
a boson.

The QCD lagrangian in eq.(\ref{LagQCD}) has a simple structure but a very rich dynamical content. It gives rise to a
complex spectrum of hadrons, it implies the striking properties of confinement and asymptotic freedom, is endowed with
an approximate chiral symmetry which is spontaneously broken, has a highly non trivial topological vacuum structure
(instantons, $U(1)_A$ symmetry breaking, strong CP violation (?)...), an intriguing phase transition diagram (colour
deconfinement, quark-gluon plasma, chiral symmetry restauration, colour superconductivity, ...).

Confinement is the property that no isolated coloured charge can exist but only colour singlet particles. For example,
the potential between a quark and an antiquark has been studied on the lattice and it has a Coulomb part at short
distances and a linearly rising term at long distances:
\beq
V_{q\bar q}~\approx~C_F[\frac{\alpha_s(r)}{r}~+....+\sigma r]\label{V}\\
\eeq
where
\beq
{\bf 1}_3~C_F~=~\sum_At^At^A~=~\frac{N_C^2-1}{2N_C}\label{CF}~{\bf 1}_3
\eeq
with $N_C$ the number of colours ($N_C=3$ in QCD). The scale dependence of $\alpha_s$ (the distance r is
Fourier-conjugated to momentum transfer) will be explained in detail in this course. The linearly rising term makes it
energetically impossible to separate a $q-\bar q$ pair. If the pair is created at one space-time point, for example in
$e^+e^-$ annihilation, and then the quark and the antiquark start moving away from each other in the center of mass frame,
it soon becomes energetically favourable to create additional pairs, smoothly distributed in rapidity between the two leading
charges, which neutralise colour and allow the final state to be reorganised into two jets of colourless hadrons, that
communicate in the central region by a number of "wee" hadrons with small energy. It is just like the familiar example of
the broken magnet: if you try to isolate a magnetic pole by stretching a dipole, the magnet breaks down and two new poles
appear at the breaking point. Very often in QCD one computes inclusive rates for partons (the fields in the lagrangian, that
is, in QCD, quarks and gluons) and takes them as equal to rates for hadrons. Partons and hadrons are considered as two
equivalent sets of complete states. This is called "global duality" and it is rather safe in the rare instance of a totally
inclusive final state. It is less so for distributions, like distributions in the invariant mass M ("local duality") where
it can be reliable only if smeared over a sufficiently wide bin in M.

Confinement is essential to explain why nuclear forces have very short range while massless gluon exchange would be long
range. Nucleons are colour singlets and they cannot exchange colour octet gluons but only colourless states. The lightest
colour singlet hadronic particles are pions. So the range of nuclear forces is fixed by the pion mass $r\simeq
m_{\pi}^{-1} \simeq 10^{-13}~cm$ : $V\approx \exp(-m_{\pi}r)/r$.

Why $SU(N_C=3)_{colour}$? The selection of $SU(3)$ as colour gauge group is unique in view of a number of constraints.
(a) The group must admit complex representations because it must be able to distinguish a quark from an antiquark. In
fact there are meson states made up of  $q\bar q$ but not analogous $qq$ bound states. Among simple groups this
restricts the choice to $SU(N)$ with $N\geq 3$, $SO(4N+2)$ with $N\geq 2$ (taking into account that $SO(6)$ has the
same algebra as $SU(4)$) and $E(6)$. (b) The group must admit a completely antisymmetric colour singlet baryon made up of 3
quarks:
$qqq$. From the study of hadron spectroscopy we know that the low lying baryons, completing an octet and a decuplet of
(flavour)
$SU(3)$ (the approximate symmetry that rotate the 3 light quarks u, d and s), are made up of three quarks and are colour
singlets. The
$qqq$ wave function must be completely antisymmetric in colour in order to agree with Fermi statistics. Indeed if we
consider, for example, a
$N^{*++}$ with spin z-component +3/2, this is made up of $(u\Uparrow u\Uparrow u\Uparrow)$ in an s-state. Thus its wave
function is totally symmetric in space, spin and flavour so that complete antisymmetry in colour is required by Fermi
statistics. In QCD this requirement is very simply satisfied by $\epsilon_{abc}q^aq^bq^c$ where a, b, c are
$SU(3)_{colour}$ indices. (c) The choice of
$SU(N_C=3)_{colour}$ is confirmed by many processes that directly measure $N_C$. Some examples are listed here. The total
rate for hadronic production in
$e^+e^-$ annihilation is linear in $N_C$. Precisely if we consider $R=\sigma(e^+e^-\rightarrow
hadrons)/\sigma_{point}(e^+e^-\rightarrow \mu^+\mu^-)$ above $b \bar b$ threshold and below $m_Z$ and we neglect small
computable radiative corrections (that will be discussed later) we have a sum of individual contributions (proportional
to $Q^2$) from
$q\bar q$ final states with $q=u,~c,~d,~s,~b$:
\beq
R~\approx~N_C [ 2 \cdot \frac{4}{9}~+~3\cdot \frac{1}{9}]~\approx~N_C \frac{11}{9}\label{R}\\  
\eeq 
The data neatly indicate $N_C=3$ as seen from Fig.~2 \cite{pdg}.  The slight excess of the data with respect to the value 11/3 is due to the QCD radiative corrections. Similarly we can consider the branching ratio
$B(W^-\rightarrow e^-\bar{\nu})$, again in Born approximation. The possible fermion-antifermion ($f\bar f$) final states are
for $f= e^-,~\mu^-,~\tau^-, d, s$ (there is no $f=b$ because the top quark is too heavy for $b\bar t$ to occur). Each channel
gives the same contribution, except that for quarks we have $N_C$ colours:

\begin{figure}[h]
\includegraphics[width=15cm]{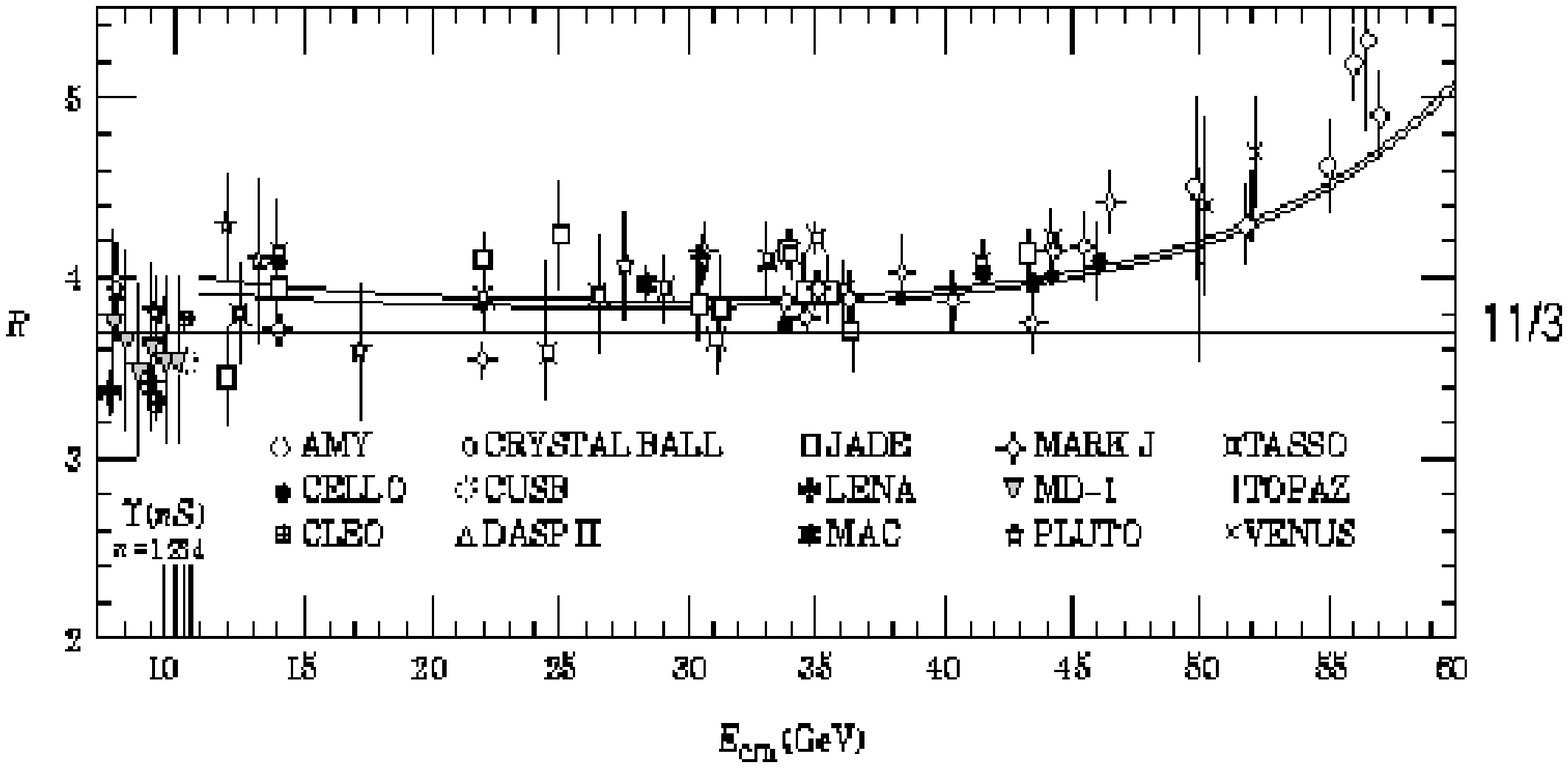}
\caption{}
\label{fig2}
\end{figure}

\beq
B(W^-\rightarrow e^-\bar{\nu})~\approx~\frac{1}{3+2N_C}\label{BW}\\
\eeq
For $N_C=3$ we obtain $B=11\%$ and the experimental number is $B=10.7\%$. Another analogous example is the branching ratio
$B(\tau^-\rightarrow
e^-\bar{\nu_e}\nu_{\tau})$. From the final state channels with $f=
e^-,~\mu^-,~d$ we find
\beq
B(\tau^-\rightarrow e^-\bar{\nu_e}\nu_{\tau})~\approx~\frac{1}{2+N_C}\label{Btau}\\
\eeq
For $N_C=3$ we obtain $B=20\%$ and the experimental number is $B=18\%$ (the less accuracy in this case is explained by
the larger radiative and phase-space corrections because the mass of $\tau^-$ is much smaller than $m_W$). An important
process that is quadratic in $N_C$ is the rate $\Gamma(\pi^0\rightarrow 2\gamma)$. This rate can be reliably calculated
from a solid theorem in field theory which has to do with the chiral anomaly:
\beq 
\Gamma(\pi^0\rightarrow 2\gamma)\approx~(\frac{N_C}{3})^2\frac{\alpha^2m_{\pi^0}^3}{32\pi^3f_{\pi}^2}~=~
(7.73\pm0.04)(\frac{N_C}{3})^2~eV\label{pi0}\\
\eeq
where the prediction is obtained for $f_{\pi}=(130.7\pm0.37)~MeV$. The experimental result is $\Gamma~=~(7.7\pm0.5)~eV$ in
remarkable agreement with $N_C=3$. There are many more experimental confirmations that $N_C=3$: for example the rate for
Drell-Yan processes (see section 5.4 ) is inversely proportional to $N_C$.

How do we get testable predictions from QCD? On the one hand there are non perturbative methods. The most important
at present is the technique of lattice simulations: it is based on first principles, it has produced very valuable
results on confinement, phase transitions, bound states, hadronic matrix elements and so on, and it is by now an
established basic tool. The main limitation is computing power and therefore there is continuous progress and a lot
of good perspectives for the future. Another class of approaches is based on effective lagrangians which provide
simpler approximations than the full theory, valid in some definite domain of physical conditions. Chiral
lagrangians are based on soft pion theorems and are valid for suitable processes at energies below $1~GeV$. Heavy
quark effective theories are obtained from expanding in inverse powers of the heavy quark mass and are
mainly important for the study of b and, to less accuracy, c decays. The approach of QCD sum rules has led to
interesting results but appears to offer not much potential for further development. Similarly specific potential
models for quarkonium have a limited range of application. On the other hand, the perturbative approach, based on
asymptotic freedom, still remains the main quantitative connection to experiment, due to its wide range of
applicability to all sorts of "hard" processes. Perturbative QCD will be the main subject in the following. I will
discuss its foundations and main applications in the next sections.

\section{Massless QCD and Scale Invariance}

The QCD lagrangian in eq.(\ref{LagQCD}) only specifies the theory at the classical level. The procedure for quantisation of
gauge theories involves a number of complications that arise from the fact that not all degrees of freedom of gauge
fields are physical because of the constraints from gauge invariance which can be used to eliminate the dependent
variables. This is already true for abelian theories and we are familiar with the QED case. One introduces a gauge fixing
term (an additional term in the lagrangian density that acts as a Lagrange multiplier in the action extremisation).
One can choose to preserve manifest Lorentz invariance. In this case, one adopts a covariant gauge, like the Lorentz
gauge, and  in QED one proceeds according to the formalism of Gupta-Bleuler. Or one can give up explicit formal covariance
and work in a non covariant gauge, like the Coulomb or the axial gauges, and only quantise the physical degrees of freedom
(the transverse components of the photon field). While this is all for an abelian gauge theory, in the non-abelian case
some additional complications arise, in particular the necessity to introduce ghosts for the formulation of Feynman rules.
There are in general as many ghost fields as gauge bosons and they appear in the form of a transformation Jacobian in the
Feynman diagram functional integral. Ghosts only propagate in closed loops and their vertices with gluons can be included
as additional terms in the lagrangian density which are fixed once the gauge fixing terms and their infinitesimal gauge
transformations are specified. We skip the detailed derivation of the complete Feynman rules in a given gauge as they
appear in Fig~1. 

Once the Feynman rules are derived we have a formal perturbative expansion but loop diagrams generate infinities. First a
regularisation must be introduced, compatible with gauge symmetry and Lorentz invariance. This is possible in QCD. In
principle one can introduce a cut-off $\Lambda$ (with dimensions of energy), for example, a' la Pauli-Villars. But at
present the universally adopted regularisation procedure is dimensional regularisation that we will briefly describe later
on. After regularisation the next step is renormalisation. In a renormalisable theory (like for all gauge theories in 4
spacetime dimensions and for QCD in particular) the dependence on the cutoff can be completely reabsorbed in a
redefinition of particle masses, of gauge coupling(s) and of wave function normalisations. After renormalisation is
achieved the perturbative definition of the quantum theory that corresponds to a classical lagrangian
like in eq.(\ref{LagQCD}) is completed. 

In the QCD Lagrangian of eq.(\ref{LagQCD}) quark masses are the only parameters with physical dimensions (we work in the
natural system of units $\hbar=c=1$). Naively we would expect that massless QCD is scale invariant. This is actually true
at the classical level. Scale invariance implies that dimensionless observables should not depend on the absolute scale
of energy but only on ratios of energy-dimensional variables. The massless limit should be relevant for the asymptotic
large energy limit of processes which are non singular for $m\rightarrow 0$.

The naive expectation that massless QCD should be scale invariant is false in the quantum theory. The scale symmetry of
the classical theory is unavoidably destroyed by the regularisation and renormalisation procedure which introduce a
dimensional parameter in the quantum version of the theory. When a symmetry of the classical theory is necessarily
destroyed by quantisation, regularisation and renormalisation one talks of an "anomaly". So, in this sense, scale
invariance in massless QCD is anomalous.

While massless QCD is finally not scale invariant, the departures from scaling are asymptotically small, logarithmic and
computable. In massive QCD there are additional mass corrections suppressed by powers of m over the energy scale (for non
singular processes in the limit $m\rightarrow 0$). At the parton level (q and g) we can conceive to apply the asymptotics
from massless QCD to processes and observables (we use the word "processes" for both) with the following properties ("hard
processes"). (a) All relevant energy variables must be large:
\beq
E_i~=~z_iQ,~~~~~~~~~Q>>m_j;~~~~~~~~~~z_i\rm{:scaling~variables~o(1)}
\label{hp}
\eeq
(b) There should be no infrared singularities (one talks of "infrared safe" processes). (c) The processes
concerned must be finite for $m\rightarrow 0$ (no mass singularities). To possibly satisfy these criteria processes must be
as "inclusive" as possible: one should include all final states with massless gluon emission and add all mass degenerate
final states (given that quarks are massless also $q-\bar q$ pairs can be massless if "collinear", that is moving together
in the same direction at the common speed of light).

Let us discuss more in detail infrared and collinear safety. Consider, for example, a quark virtual line that ends up into
a real quark plus a real gluon (Fig.~3).

\begin{figure}[h]
\begin{center}
\includegraphics[width=7cm]{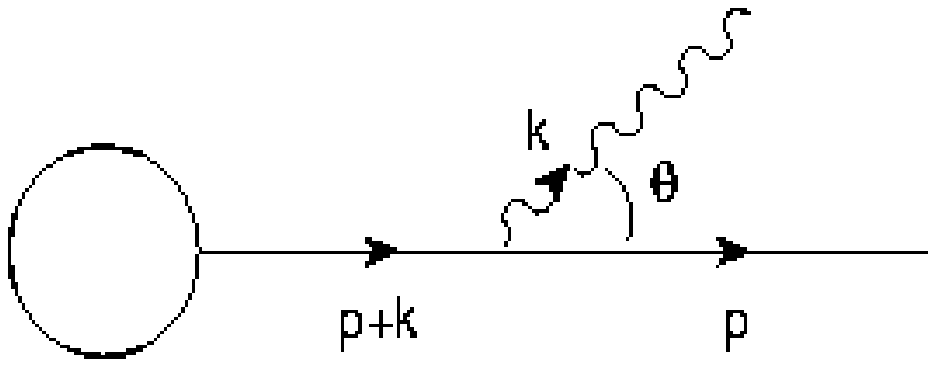}
\caption[]{}
\label{fig3}
\end{center}
\end{figure}

For the propagator we have:
\beq
\rm{propagator}~=~\frac{1}{(p+k)^2-m^2}~=~\frac{1}{2(p\cdot k)}~=~\frac{1}{2E_kE_p}\cdot\frac{1}{1-\beta_p\cos{\theta}}
\label{prop}\\
\eeq 
Since the gluon is massless, $E_k$ can vanish and this corresponds to an infrared singularity. Remember that we have to
take the square of the amplitude and integrate over the final state phase space, or all together, $dE_k/E_k$. So we get $1/E_k^2$ from the squared
amplitude and $d^3k/E_k\sim E_kdE_k$ from the phase space. Also, for
$m\rightarrow 0$,
$\beta_p=\sqrt{1-m^2/E_p^2}\rightarrow 1$ and $(1-\beta_p\cos{\theta})$ vanishes at $\cos{\theta}=1$. This leads to a
collinear mass singularity.

There are two very important theorems on infrared and mass singularities. The first one is the Bloch-Nordsieck theorem:
infrared singularities cancel between real and virtual diagrams (see Fig.~4) when all resolution indistinguishable final
states are added up. For example, for each real detector there is a minimum energy of gluon radiation that can be
detected. For the cancellation of infrared divergences, one should add all possible gluon emission with a total energy
below the detectable minimum. The second one is the Kinoshita-Lee-Nauenberg theorem: mass singularities connected with an
external particle of mass m are canceled if all degenerate states (that is with the same mass) are summed up. That is for
a final state particle of mass m we should add all final states that in the limit
$m\rightarrow 0$ have the same mass, also including gluons and massless pairs. If a completely inclusive final state is
taken, only the mass singularities from the initial state particles remain (we shall see that they will be absorbed inside
the non perturbative parton densities, which are probability densities of finding the given parton in the initial hadron).

\begin{figure}[h]
\begin{center}
\includegraphics[width=7cm]{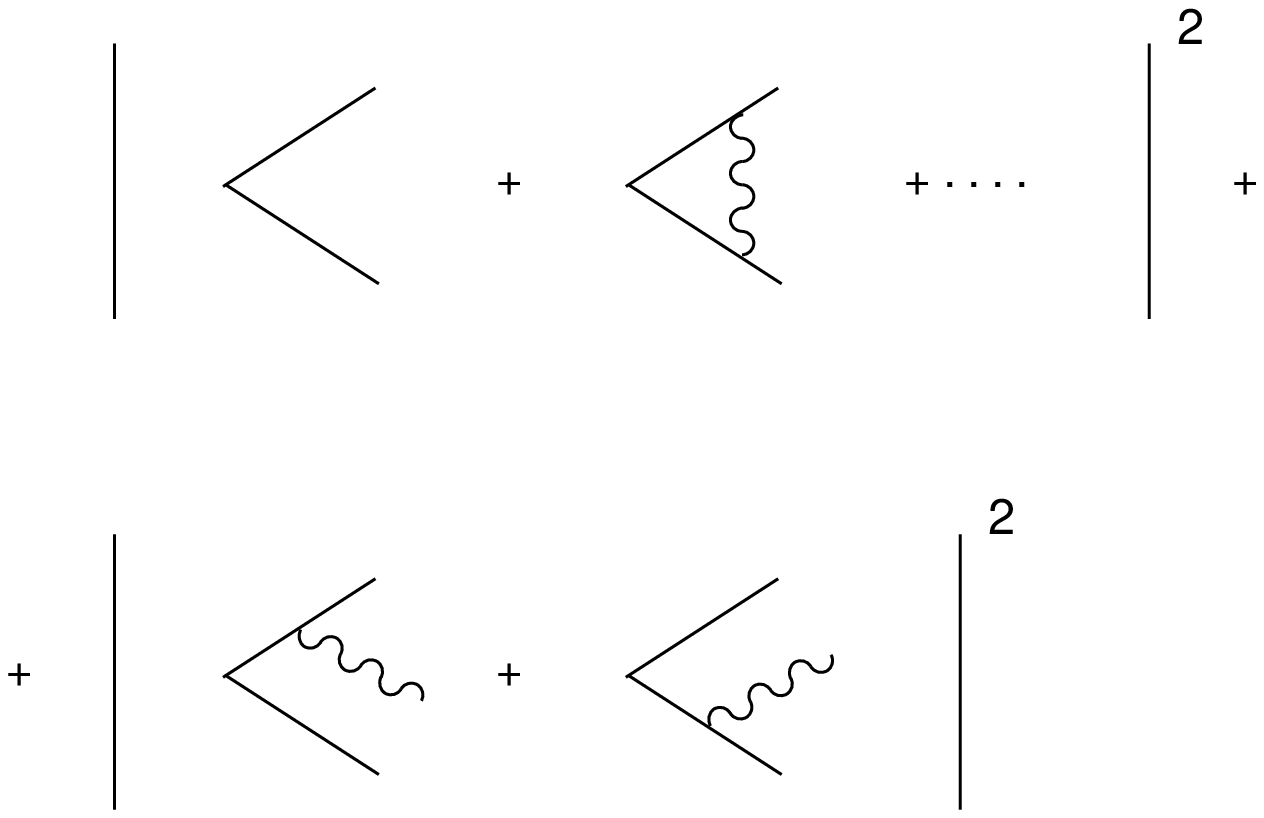}
\caption[]{}
\label{fig4}
\end{center}
\end{figure}

Hard processes to which the massless QCD asymptotics can possibly apply must be infrared and collinear safe, that is must
satisfy the requirements from the Bloch-Nordsieck and the Kinoshita-Lee-Nauenberg theorems. We give now some examples of
important hard processes. One of the simplest hard processes is the totally inclusive cross section for hadron production in
$e^+e^-$ annihilation, Fig.~5, parameterised in terms of the already mentioned dimensionless observable
$R=\sigma(e^+e^-\rightarrow hadrons)/\sigma_{point}(e^+e^-\rightarrow \mu^+\mu^-)$. The pointlike cross section in the
denominator is given by
$\sigma_{point} = 4\pi\alpha^2/3s$, where $s=Q^2=4E^2$ is the squared total center of mass energy and $Q$ is the mass of
the exchanged virtual gauge boson. At parton level the final state is $(q\bar q~+~n~g~+~n'~q'\bar q')$ and n and n' are
limited at each order of perturbation theory. It is assumed that the conversion of partons into hadrons does not affect
the rate (it happens with probability 1). We have already mentioned that in order for this to be true within a given
accuracy an averaging over a sufficiently large bin of $Q$ must be understood. The binning width is larger in the
vicinity of thresholds: for example when one goes across the charm $c\bar c$ threshold the physical cross-section shows
resonance bumps which are absent in the smooth partonic counterpart which however gives an average of the cross-section.

\begin{figure}[h]
\begin{center}
\includegraphics[width=7cm]{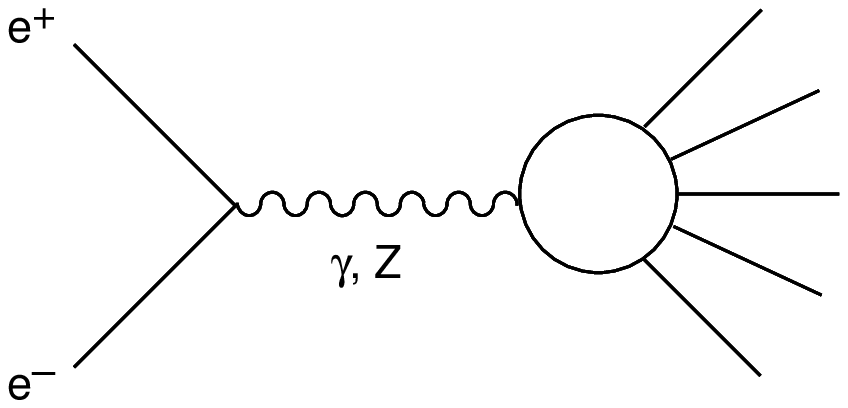}
\caption[]{}
\label{fig5}
\end{center}
\end{figure}

A very important class of hard processes is Deep Inelastic Scattering (DIS)
\beq 
l~+~N\rightarrow l'~+~X~~~~~~~~~~~~l=e^{\pm}, \mu^{\pm}, \nu, \bar{\nu}\label{DIS}
\eeq 
which has played and still plays a very important role for our understanding of QCD and of nucleon structure. 
For the processes in eq.(\ref{DIS}), shown in Fig~6, we have, in the lab system where the nucleon of mass m is
at rest:
\beq Q^2~=~-q^2~=~-(k-k')^2~=~4EE'\sin^2{\theta/2};~~~~~~~~m\nu~=~(p.q);~~~~~~~~x~=~\frac{Q^2}{2m\nu}\label{kin}  
\eeq
In this case the virtual momentum $q$ of the gauge boson is spacelike. $x$ is the familiar Bjorken variable.

\begin{figure}[h]
\begin{center}
\includegraphics[width=6cm]{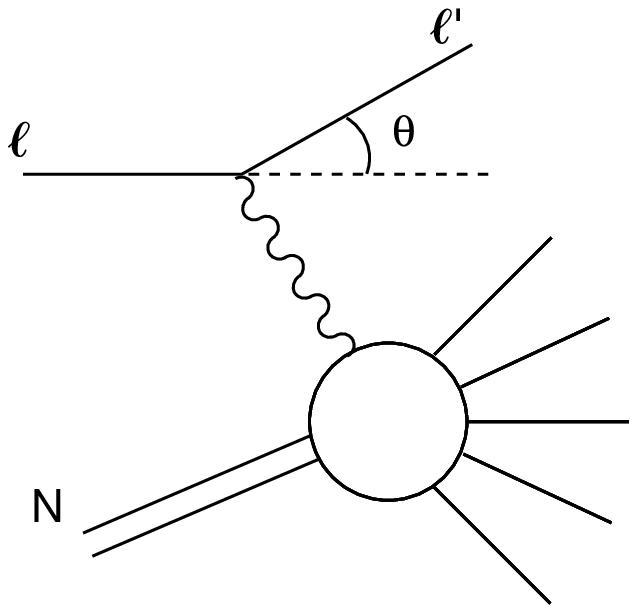}
\caption[]{}
\label{fig6}
\end{center}
\end{figure}

\section{The Renormalisation Group and Asymptotic Freedom}

In this section we aim at providing a reasonably detailed introduction to the renormalisation group formalism and the
concept of running coupling which leads to the result that QCD has the property of asymptotic freedom. We start with a
summary on how renormalisation works. 

In the simplest conceptual situation imagine that we implement regularisation of divergent integrals by introducing a
dimensional cut-off $\Lambda$ that respects gauge and Lorentz invariance. The dependence of renormalised quantities on
$\Lambda$ is eliminated by absorbing it into a redefinition of m (the quark mass: for simplicity we assume a single
flavour here), the gauge coupling $e$ (can be $e$ in QED or $e_s$ in QCD) and the wave function renormalisation factors
$Z^{1/2}_{q,g}$ for q and g, using suitable renormalisation conditions (that is precise definitions of m, g and Z). For
example we can define the renormalised mass m as the position of the pole in the quark propagator and, similarly, the
normalisation $Z_q$ as the residue at the pole:
\beq
\rm{Propagator}~=~\frac{Z_q}{p^2-m^2}~+~\rm{no-pole~terms}\label{mZ}\\
\eeq
The renormalised coupling $e$ can be defined in terms of a renormalised 3-point vertex at some specified values of the
external momenta. We now become more specific by concentrating in the case of massless QCD. If we start from a vanishing
mass at the classical (or "bare") level, $m_0=0$, the mass is not renormalised because it is protected by a symmetry,
chiral symmetry. The conserved currents of chiral symmetry are axial currents: $\bar q\gamma_{\mu}\gamma_5q$. The
divergence of the axial current gives, by using the Dirac equation, $\partial^{\mu}(\bar q\gamma_{\mu}\gamma_5q)~=~2m\bar
q\gamma_5q$. So the axial current  and corresponding axial charge are conserved in the massless limit. Since QCD 
is a vector theory we have not to worry about chiral anomalies in this respect. So one can choose a regularisation that
preserves chiral symmetry besides gauge and Lorentz symmetry. Then the renormalised mass remains zero. The renormalised
propagator has the form in eq.(\ref{mZ}) with $m=0$. 

The renormalised coupling $e_s$ can be defined from the
renormalised 3-gluon vertex at a scale $-\mu^2$ (Fig.~7):
\beq
V_{bare}(p^2,q^2,r^2)~=~ZV_{ren}(p^2,q^2,r^2),~~~~~~Z=Z_g^{3/2}Z_V,~~~~~~e_s=V_{ren}(-\mu^2,-\mu^2,-\mu^2)\label{e}\\
\eeq

\begin{figure}[h]
\begin{center}
\includegraphics[width=11cm]{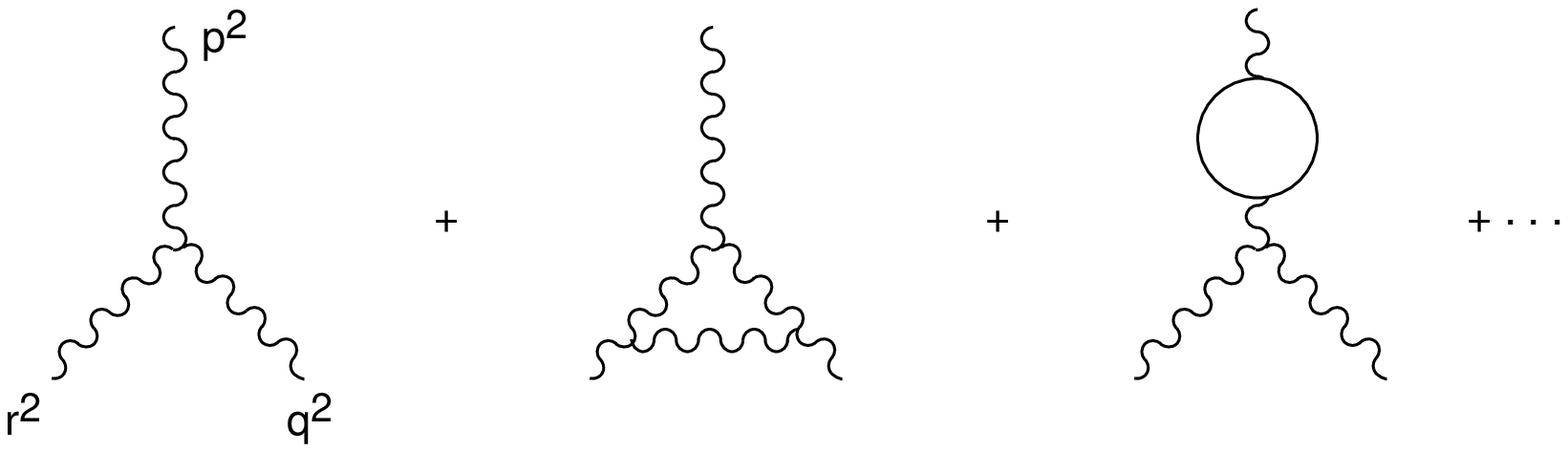}
\caption[]{}
\label{fig7}
\end{center}
\end{figure}

Here $V_{bare}$ is what is obtained from computing the Feynman diagrams including, for example, the 1-loop corrections at 
the lowest non
trivial order ($V_{bare}$ is defined so that it coincides with $e_{s0}$ in lowest order). It contains the
cut-off
$\Lambda$ but does not know
$\mu$. $Z$ is a factor that depends on the cut-off. Because of infrared singularities the defining scale
$\mu$ cannot vanish. The negative value
$-\mu^2<0$ is chosen to stay away from physical cuts (a gluon with negative virtual mass cannot decay). Similarly 
we can define $Z_g$ from the massless gluon propagator at the same scale $-\mu^2$ (the
vanishing mass of the gluon is guaranteed by gauge invariance). 

After computing
all 1-loop diagrams in Fig.~7 we have:
\bea
V_{bare}(p^2,p^2,p^2)~&=&~e_{0s}[1+c\alpha_{0s}\cdot \log{\frac{\Lambda^2}{p^2}~+...]}~=~\nonumber\\
&&~=~[1+c\alpha_{s}\cdot
\log{\frac{\Lambda^2}{-\mu^2}~+~...]}e_s[1+c\alpha_{s}\cdot
\log{\frac{-\mu^2}{p^2}~+...]}~=~ZV_{ren}
\label{Vren}
\eea
Note the replacement of $e_0$ with $e$ in the second step. The definition of $e_s$ demands that one precisely
specifies what is included in $Z$. For this, in a given renormalisation scheme, a prescription is
fixed to specify the finite terms that go into Z (i.e. the terms of order $\alpha_s$ that accompany $\log{\Lambda^2}$).
Then
$V_{ren}$ is specified and the renormalised coupling is defined from it according to eq.(\ref{e}). For example, in the
momentum subtraction scheme we define $V_{ren}(p^2,p^2,p^2)=e_s~+~V_{bare}(p^2,p^2,p^2)-V_{bare}(-\mu^2,-\mu^2,-\mu^2)$,
which is equivalent to say, at 1-loop, that all finite terms that do not vanish at $p^2=-\mu^2$ are included in Z.

A crucial observation is that $V_{bare}$ depends on $\Lambda$ but not on $\mu$, which is only introduced when Z,
$V_{ren}$ and hence $\alpha_{s}$ are defined. (From here on, for shorthand, we write $\alpha$ to indicate either the QED
coupling or the QCD coupling $\alpha_{s}$). More in general for a generic Green function G, we similarly have:
\beq
G_{bare}(\Lambda^2,\alpha_0,p_i^2)~=~ZG_{ren}(\mu^2,\alpha,p_i^2)\label{G}\\
\eeq
so that we have:
\beq
\frac{dG_{bare}}{d\log{\mu^2}}~=~\frac{d}{d\log{\mu^2}}[ZG_{ren}]~=~0\label{dG}\\
\eeq
or
\beq
Z[\frac{\partial}{\partial\log{\mu^2}}~+~\frac{\partial
\alpha}{\partial\log{\mu^2}}\frac{\partial}{\partial \alpha}~+~
\frac{1}{Z}\frac{\partial
Z}{\partial\log{\mu^2}}]G_{ren}~=~0\label{RGE1}\\
\eeq
Finally the renormalisation group equation (RGE) can be written as:
\beq
[\frac{\partial}{\partial\log{\mu^2}}~+~\beta(\alpha)\frac{\partial}{\partial \alpha}~+~
\gamma(\alpha)]G_{ren}~=~0\label{RGE2}\\
\eeq
where
\beq
\beta(\alpha)~=~\frac{\partial\alpha}{\partial\log{\mu^2}}\label{beta}\\
\eeq
and
\beq
\gamma(\alpha)~=~\frac{\partial\log{Z}}{\partial\log{\mu^2}}\label{gamma}\\
\eeq
Note that $\beta(\alpha)$ does not depend on which Green function $G$ we are considering, but it is a property of the
theory and the renormalisation scheme adopted, while $\gamma(\alpha)$ also depends on $G$. 

Assume that we want to apply the RGE to some hard process at a large scale Q, related to a Green function G that we can
always take as adimensional (by multiplication by a suitable power of Q). Since the interesting dependence on Q 
will be logarithmic we introduce the variable t as :
\beq
t~=~\log{\frac{Q^2}{\mu^2}}\label{t}\\
\eeq
Then we can write $G_{ren}\equiv F(t,\alpha,x_i)$ where $x_i$ are scaling variables (we often omit to write them in the
following). In the naive scaling limit
$F$ should be independent of $t$. To find the actual dependence on $t$, we want to solve the RGE
\beq
[-\frac{\partial}{\partial t}~+~\beta(\alpha)\frac{\partial}{\partial \alpha}~+~
\gamma(\alpha)]G_{ren}~=~0\label{RGE3}\\
\eeq
with a given boundary condition at $t=0$ (or $Q^2=\mu^2$): $F(0,\alpha)$. 

We first solve the RGE in the simplest case that $\gamma(\alpha)=0$. This is not an unphysical case: for example, it
applies to $R_{e^+e^-}$ where the vanishing of $\gamma$ is related to the non renormalisation of the electric charge in
QCD (otherwise the proton and the electron charge would not exactly compensate: this will be better explained later).
So we consider the equation:
\beq
[-\frac{\partial}{\partial t}~+~\beta(\alpha)\frac{\partial}{\partial \alpha}]G_{ren}~=~0\label{RGE4}\\
\eeq
The solution is simply
\beq
F(t,\alpha)~=~F[0,\alpha(t)]\label{Fsol1}\\
\eeq
where the "running coupling" $\alpha(t)$ is defined by:
\beq
t~=~\int_{\alpha}^{\alpha(t)}\frac{1}{\beta(\alpha')}d\alpha'\label{run}\\
\eeq
Note that from this definition it follows that $\alpha(0)~=~\alpha$, so that the boundary condition is also satisfied.
To prove that $F[0,\alpha(t)]$ is indeed the solution, we first take derivatives with respect of t and $\alpha$ (the two
independent variables) of both sides of eq.(\ref{run}). By taking $d/dt$ we obtain
\beq
1~=~\frac{1}{\beta(\alpha(t)}\frac{\partial\alpha(t)}{\partial t}\label{ddt}\\
\eeq
We then take $d/d\alpha$ and obtain
\beq
0~=~-\frac{1}{\beta(\alpha)}~+~\frac{1}{\beta(\alpha(t)}\frac{\partial\alpha(t)}{\partial \alpha}\label{dda}\\
\eeq
These two relations make explicit the dependence of the running coupling on t and $\alpha$:
\bea
\frac{\partial\alpha(t)}{\partial t}~=~\beta(\alpha(t))\label{runt}\\
\frac{\partial\alpha(t)}{\partial \alpha}~=~\frac{\beta(\alpha(t))}{\beta(\alpha)}\label{runa}\\
\nonumber
\eea
Using these two equations one immediately checks that $F[0,\alpha(t)]$ is indeed the solution.

Similarly, one finds that the solution of the more general equation with $\gamma\not=0$, eq.(\ref{RGE3}), is given by:
\beq
F(t,\alpha)~=~F[0,\alpha(t)]\exp{\int_{\alpha}^{\alpha(t)}\frac{\gamma(\alpha')}{\beta(\alpha')}d\alpha'}\label{Fsol2}\\
\eeq
In fact the sum of the two derivatives acting on the factor $F[0,\alpha(t)]$ vanishes and the exponential is by itself a
solution of the complete equation. Note that the boundary condition is also satisfied.

The important point is the appearance of the running coupling that determines the asymptotic departures from scaling. The
next step is to study the functional form of the running coupling. From eq.(\ref{runt}) we see that the rate of change
with t of the running coupling is determined by the $\beta$ function. In turn $\beta(\alpha)$ is determined by the $\mu$
dependence of the renormalised coupling through eq.(\ref{beta}). Clearly there is no dependence on $\mu$ of the basic
3-gluon vertex in lowest order (order $e$). The dependence starts at 1-loop, that is at order $e^3$ (one extra gluon has
to be emitted and reabsorbed). Thus we obtain that in perturbation theory:
\beq
\frac{\partial e}{\partial \log{\mu^2}}~\propto~e^3\label{de}\\
\eeq
Recalling that $\alpha~=~e^2/4\pi$, we have:
\beq
\frac{\partial \alpha}{\partial \log{\mu^2}}~\propto~2e\frac{\partial e}{\partial \log{\mu^2}}~\propto~e^4
~\propto \alpha^2\label{da}\\
\eeq
Thus the behaviour of $\beta(\alpha)$ in perturbation theory is as follows:
\beq
\beta(\alpha)~=~\pm b\alpha^2[1~+~b'\alpha~+...]\label{betapert}\\
\eeq
Since the sign of the leading term is crucial in the following discussion, we stipulate that always $b>0$ and we
make the sign explicit in front. By direct calculation at 1-loop one finds:
\beq
\rm{QED:}~~~~~~~~\beta(\alpha)~\sim~+b\alpha^2~+.....~~~~~~~~~~~b~=~\sum_i\frac{N_C(Q^2)_i}{3\pi}
\label{beQED}\\
\eeq
where $N_C = 3$ for quarks and $N_C = 1$ for leptons and the sum runs over all fermions of charge $Qe$ that are coupled. Also, one finds:
\beq
\rm{QCD:}~~~~~~~~\beta(\alpha)~\sim~-b\alpha^2~+.....~~~~~~~~~~~b~=~\frac{11N_C-2n_f}{12\pi}\label{beQCD}\\
\eeq
where, as usual, $n_f$ is the number of coupled flavours of quarks (we assume here that $n_f~\le~16$ so that $b>0$ in QCD).
If
$\alpha(t)$ is small we can compute
$\beta(\alpha(t))$ in perturbation theory. The sign in front of $b$ then decides the slope of the coupling: $\alpha(t)$
increases with t (or
$Q^2$) if $\beta$ is positive at small $\alpha$ (QED), or $\alpha(t)$ decreases with t (or
$Q^2$) if $\beta$ is negative at small $\alpha$ (QCD). A theory like QCD where the running coupling vanishes
asymptotically at large $Q^2$ is called (ultraviolet) "asymptotically free". An important result that can be proven is
that in 4 spacetime dimensions all and only non-abelian gauge theories can be asymptotically free.

Going back to eq.(\ref{run}) we replace $\beta(\alpha)~\sim~\pm b\alpha^2$, do the integral and perform a simple algebra.
We find 
\beq
\rm{QED:}~~~~~~~~\alpha(t)~\sim~\frac{\alpha}{1-b\alpha t}\label{beQED1}\\
\eeq
and
\beq
\rm{QCD:}~~~~~~~~\alpha(t)~\sim~\frac{\alpha}{1+b\alpha t}\label{beQCD1}\\
\eeq
A slightly different form is often used in QCD. Defining $1/\alpha~=~b\log{\mu^2/\Lambda_{QCD}^2}$ we can write:
\beq
\alpha(t)~\sim~\frac{1}{\frac{1}{\alpha}~+~bt}~=~\frac{1}{b\log{\frac{\mu^2}{\Lambda_{QCD}^2}}~+~b\log{\frac{Q^2}{\mu^2}}}
~=~\frac{1}{b\log{\frac{Q^2}{\Lambda_{QCD}^2}}}\label{alfaQCD}\\
\eeq
We see that $\alpha(t)$ decreases logarithmically with $Q^2$ and that one can introduce a dimensional parameter
$\Lambda_{QCD}$ that replaces $\mu$. Often in the following we will simply write $\Lambda$ for $\Lambda_{QCD}$, assuming
that confusion with $\Lambda~=~$ ultraviolet~cut-off is avoided by the reader. Note that it is clear that $\Lambda$
depends on the particular definition of $\alpha$, not only on the defining scale $\mu$ but also on the renormalisation
scheme (see, for example, the discussion in the next session). Through the parameter $b$, and in general through the
$\beta$ function, it also depends on the number $n_f$ of coupled flavours. It is very important to note that QED and QCD
are theories with "decoupling":  up to the scale $Q$ only quarks with masses $m<<Q$
contribute to the running of $\alpha$. This is clearly very important, given that
all applications of perturbative QCD so far apply to energies below the top quark mass $m_t$. For the validity of the
decoupling theorem it is necessary that the theory where all the heavy particle internal lines are eliminated is still
renormalisable and that the coupling constants do not vary with the mass. These requirements are true for the mass of heavy
quarks in QED and QCD, but are not true in the electroweak theory where the elimination of the top would violate $SU(2)$
symmetry (because the t and b left quarks are in a doublet) and the quark couplings to the Higgs multiplet (hence to the
longitudinal gauge bosons) are proportional to the mass. In conclusion, in QED and QCD, quarks with $m>>Q$ do not contribute
to $n_f$ in the coefficients of the relevant $\beta$ function. The effects of heavy quarks are power suppressed and can be
taken separately into account. For example, in $e^+e^-$ annihilation for
$2m_c<Q<2m_b$ the relevant asymptotics is for $n_f=4$, while for $2m_b<Q<2m_t$ $n_f=5$. Going accross the $b$ threshold
the $\beta$ function coefficients change, so the $\alpha(t)$ slope changes. But $\alpha(t)$ is continuous, so that
$\Lambda$ changes so as to keep constant $\alpha(t)$ at the matching point at $Q\sim o(m_b)$. The effect on $\Lambda$ is
large: approximately $\Lambda_5~\sim~0.65\Lambda_4$. 

Note the presence of a pole in eqs.(\ref{beQED1},\ref{beQCD1}) at $\pm b\alpha t~=~1$, called the Landau pole, who realised
its existence in QED already in the '50's. For $\mu~\sim m_e$ (in QED) the pole occurs beyond the Planck mass. In QCD the Landau
pole is located for negative $t$ or at $Q<\mu$ in the region of light hadron masses. Clearly the issue of the definition
and the behaviour of the physical coupling in the region around the Landau pole is a problem that lies
outside the domain of perturbative QCD.

The non leading terms in the asymptotic behaviour of the running coupling can in principle be evaluated going back to
eq.(\ref{betapert}) and computing $b'$ at 2-loops and so on. But in general the perturbative coefficients of
$\beta(\alpha)$ depend on the definition of the renormalised coupling $\alpha$ (the renormalisation scheme), so one
wonders whether it is worthwhile to do a complicated calculation to get $b'$ if then it must be repeated for a
different definition or scheme. In this respect it is interesting to remark that actually both $b$ and $b'$ are independent
of the definition of $\alpha$, while higher order coefficients do depend on that. Here is the simple proof. Two different
perturbative definitions of $\alpha$ are related by $\alpha'~\sim~\alpha(1~+~c_1\alpha~+~...)$. Then we have:
\bea
\beta(\alpha')~=~\frac{d\alpha'}{d\log{\mu^2}}~&=&~\frac{d\alpha}{d\log{\mu^2}}(1~+~2c_1\alpha~+~...)\nonumber\\
&~=~&\pm b\alpha^2(1~+~b'\alpha~+~...)(1~+~2c_1\alpha~+~...)\nonumber\\
&~=~&\pm b\alpha'^2(1~+~b'\alpha'~+~...
\label{proof}
\eea
which shows that, up to the first subleading order, $\beta(\alpha')$ has the same form as $\beta(\alpha)$. 

In QCD ($N_C=3$)
one has calculated:
\beq
b'~=~\frac{153-19n_f}{2\pi(33-2n_f)}\label{b'}\\
\eeq
By taking $b'$ into account one can write the expression of the running coupling at next to the leading order (NLO):
\beq
\alpha(Q^2)~=~\alpha_{LO}(Q^2)[1~-~b'\alpha_{LO}(Q^2)\log{\log{\frac{Q^2}{\Lambda^2}}}~+~...]\label{NLOa}\\
\eeq
where $\alpha_{LO}^{-1}~=~b\log{Q^2/\Lambda^2}$ is the LO result.

Summarizing, we started from massless classical QCD which is scale invariant. But we have seen that the procedure of
quantisation, regularisation and renormalisation necessarily breaks scale invariance. In the quantum QCD theory there is a
scale of energy, $\Lambda_{QCD}$, which from experiment is of the order of a few hundred MeV, its precise value depending
on the definition, as we shall see in detail. Adimensional quantities depend on the energy scale through the running
coupling which is a logarithmic function of
$Q^2/\Lambda^2$. In QCD the running coupling decreases logarithmically at large $Q^2$ (asymptotic freedom), while in QED
the coupling has the opposite behaviour.

\section{More on the Running Coupling}

In the previous section we have introduced the renormalised coupling $\alpha$ in terms of the 3-gluon vertex at
$p^2=-\mu^2$ (momentum subtraction). The Ward identities of QCD then ensure that the coupling defined from other vertices
like the $\bar q qg$ vertex  are renormalised in the same way and the finite radiative corrections are related. But at
present the universally adopted definition of $\alpha_s$ is in terms of dimensional regularisation because of computational
simplicity which is essential given the great complexity of present day calculations. So we now briefly review the
principles of dimensional regularisation and the definition of Minimal Subtraction ($MS$) and Modified Minimal Subtraction
($\overline{MS} $). The $\overline{MS}$ definition of $\alpha_s$ is the one most commonly adopted in the literature and a
value quoted for it is nomally referring to this definition.

Dimensional Regularisation (DR) is a gauge and Lorentz invariant regularisation that consists in formulating the theory in 
$D<4$ spacetime dimensions in order to make loop integrals ultraviolet finite. In DR one rewrites the theory in D
dimensions (D is integer at the beginning, but then we will see that the expression of diagrams makes sense at all D except
for isolated singularities). The metric tensor is extended into a $D\times D$ matrix $g_{\mu\nu}~=~diag(1,-1,-1,....,-1)$ and
4-vectors are given by $k^{\mu}~=~(k^0,k^1,...,k^{D-1})$. The Dirac $\gamma^{\mu}$ are $f(D) \times f(D)$ matrices and it
is not important what is the precise form of the function $f(D)$. It is sufficient to extend the usual algebra in a
straightforward way like $\gamma^{\mu}\gamma^{\nu}\gamma_{\mu}~=~-(D-2)\gamma^{\nu}$ or
$Tr(\gamma^{\mu}\gamma^{\nu})~=~f(D)g_{\mu\nu}$. 

The physical dimensions of fields change in D dimensions and, as a
consequence, the gauge couplings become dimensional $e_D~=~\mu^\epsilon e$, where $e$ is adimensional, $D~=~4-2\epsilon$
and $\mu$ is a scale of mass (this is how a scale of mass is introduced in the DR of massless QCD!). The dimension of
fields is determined by requiring that the action $S~=~\int d^Dx {\cal L}$ is adimensional. By inserting for ${\cal L}$ terms
like
$m\bar\Psi \Psi$ or $m^2\phi^{\dagger} \phi$ or $e\bar\Psi \gamma^{\mu}\Psi A_{\mu}$ the dimensions of the fields and
coupling are determined as: $m, \Psi, \phi, A_\mu, e~=~1, (D-1)/2, (D-2)/2, (D-2)/2, (4-D)/2$, respectively. The formal
expression of loop integrals can be written for any D. For example:
\beq
\int\frac{d^Dk}{(2\pi)^D}\frac{1}{(k^2-m^2)^2}~=~\frac{\Gamma(2-D/2)(-m^2)^{D/2-2}}{(4\pi)^{D/2}}\label{int}\\
\eeq
For $D~=~4-2\epsilon$ one can expand using:
\beq
\Gamma(\epsilon)~=~\frac{1}{\epsilon}~-~\gamma_E~+~o(\epsilon),~~~~~~~~~~\gamma_E~=~0.5772.....\label{gamexp}\\
\eeq
For some Green function G, normalised to 1 in lowest order, (like V/e with V the 3-g vertex function at the symmetric
point $p^2=q^2=r^2$, considered in the previous section) we typically find at 1-loop:
\beq
G_{bare}~=~1~+~\alpha_0(\frac{-\mu^2}{p^2})^\epsilon~[c(\frac{1}{\epsilon}~+~\log{4\pi}~-~\gamma_E)~+~d~+~o(\epsilon)]\label{Gexp}\\
\eeq
In $\bar M \bar S$ one rewrites this at 1-loop accuracy (diagram by diagram: this is a virtue of the method): 
\bea
G_{bare}~=~ZG_{ren}\nonumber\\
Z~=~1~+~\alpha~[c(\frac{1}{\epsilon}~+~\log{4\pi}~-~\gamma_E)]\nonumber\\
G_{ren}~=~1~+~\alpha~[c\log{\frac{-\mu^2}{p^2}}~+~d]\nonumber\\
\label{MSbar}
\eea
In the original $MS$ prescription only $1/\epsilon$ was subtracted (that clearly plays the role of a cutoff) and not also
$\log{4\pi}$ and
$\gamma_E$. Later, since these constants  always appear from the expansion of $\Gamma$ functions it was decided to modify
$MS$ into
$\overline {MS}$. Note that the $\overline {MS}$ definition of $\alpha$ is different than that in the momentum subtraction
scheme because the finite terms (those beyond logs) are different. In particular here $\delta G_{ren}$ does not vanish at
$p^2=-\mu^2$. 

The third coefficient of the QCD $\beta$ function is also known in the $\overline{MS}$ prescription (recall that only the
first two coefficients are scheme independent). Translated in numbers, for $n_f=5$ one obtains \cite{tar}:
\beq
\beta(\alpha)~=~-0.610\alpha^2[1~+~1.261...\frac{\alpha}{\pi}~+~1.475...(\frac{\alpha}{\pi})^2~+...]\label{beta3}\\
\eeq 
It is interesting to remark that the expansion coefficients are all of order 1, so that the $\bar M \bar S$ expansion looks
well behaved. 

\section{Application to Hard processes}
\subsection{$R_{e^+e^-}$ and Related Processes}

The simplest hard process is $R_{e^+e^-}$ that we have already started to discuss. $R$ is adimensional and in perturbation
theory is given by
$R~=~N_C\sum_i Q^2_i F(t,\alpha_s)$, where $F~=~1~+~o(\alpha_S)$. We have already mentioned that for this process the
"anomalous dimension" function vanishes: $\gamma(\alpha_s)=0$ because of electric charge non renormalisation by strong
interactions. Let us review how this happens in detail. The diagrams that are relevant for charge renormalisation in QED
at 1-loop are shown in Fig.~8. The Ward identity that follows from gauge invariance in QED imposes that the vertex
($Z_V$) and the self-energy
($Z_f$) renormalisation factors cancel and the only divergence remains in $Z_\gamma$, the vacuum polarization of the
photon. So the charge is only renormalised by the photon blob, hence it is universal (the same factor for all fermions,
independent of their charge) and is not affected by QCD at 1-loop. It is true that at higher orders the photon vacuum
polarization diagram is affected by QCD (for example, at 2-loops we can exchange a gluon between the quarks in the photon
loop) but the renormalisation induced by the vacuum polarisation diagram remains independent of the nature of the fermion to
which the photon line is attached. The gluon contributions to the vertex
($Z_V$) and to the self-energy
($Z_f$) cancel because they have exactly the same structure as in QED, and there is no gluon contribution to the lowest
order photon vacuum polarisation blob. So $\gamma(\alpha_s)=0$. 

\begin{figure}[h]
\begin{center}
\includegraphics[width=11cm]{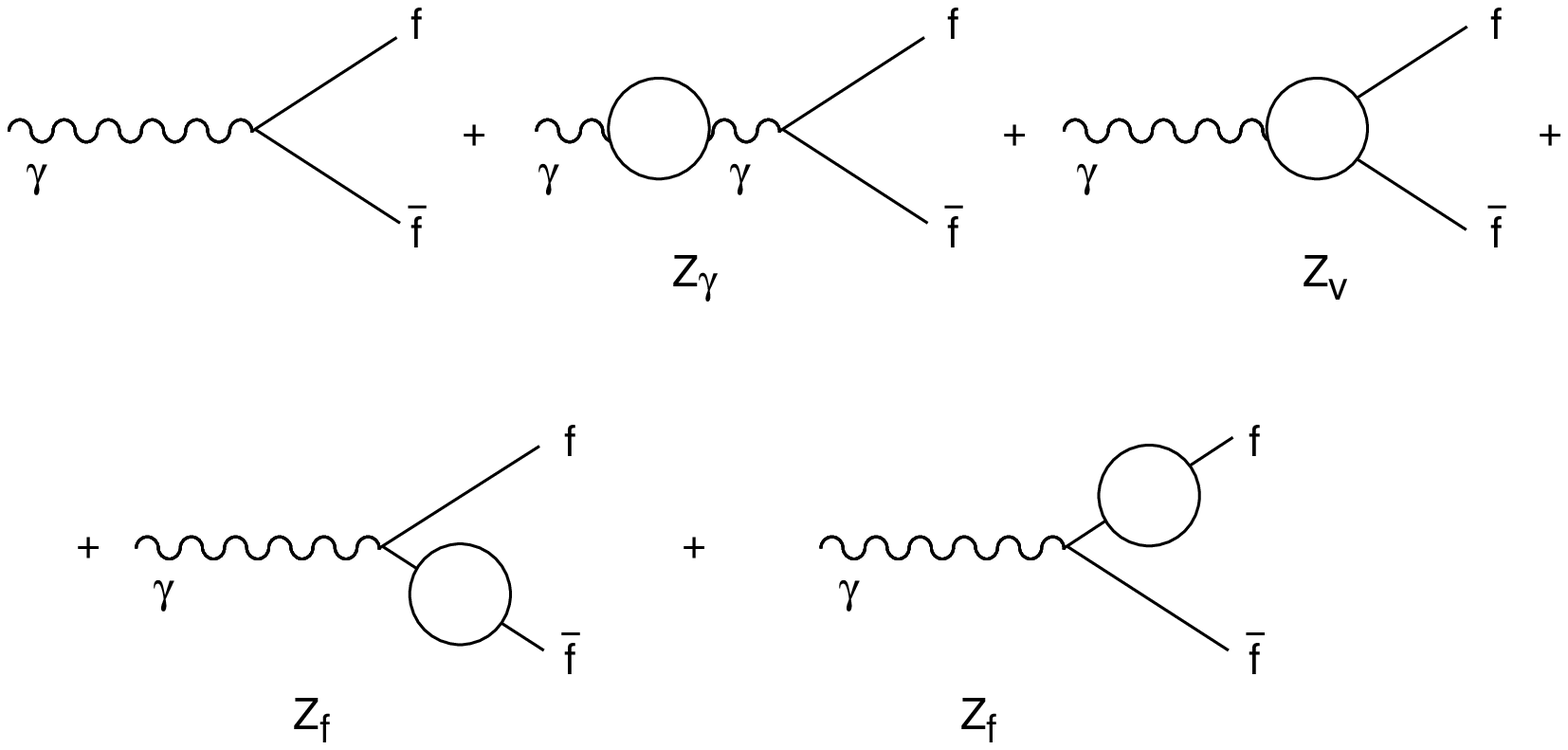}
\caption[]{}
\label{fig8}
\end{center}
\end{figure}

\begin{figure}[h]
\begin{center}
\includegraphics[width=11cm]{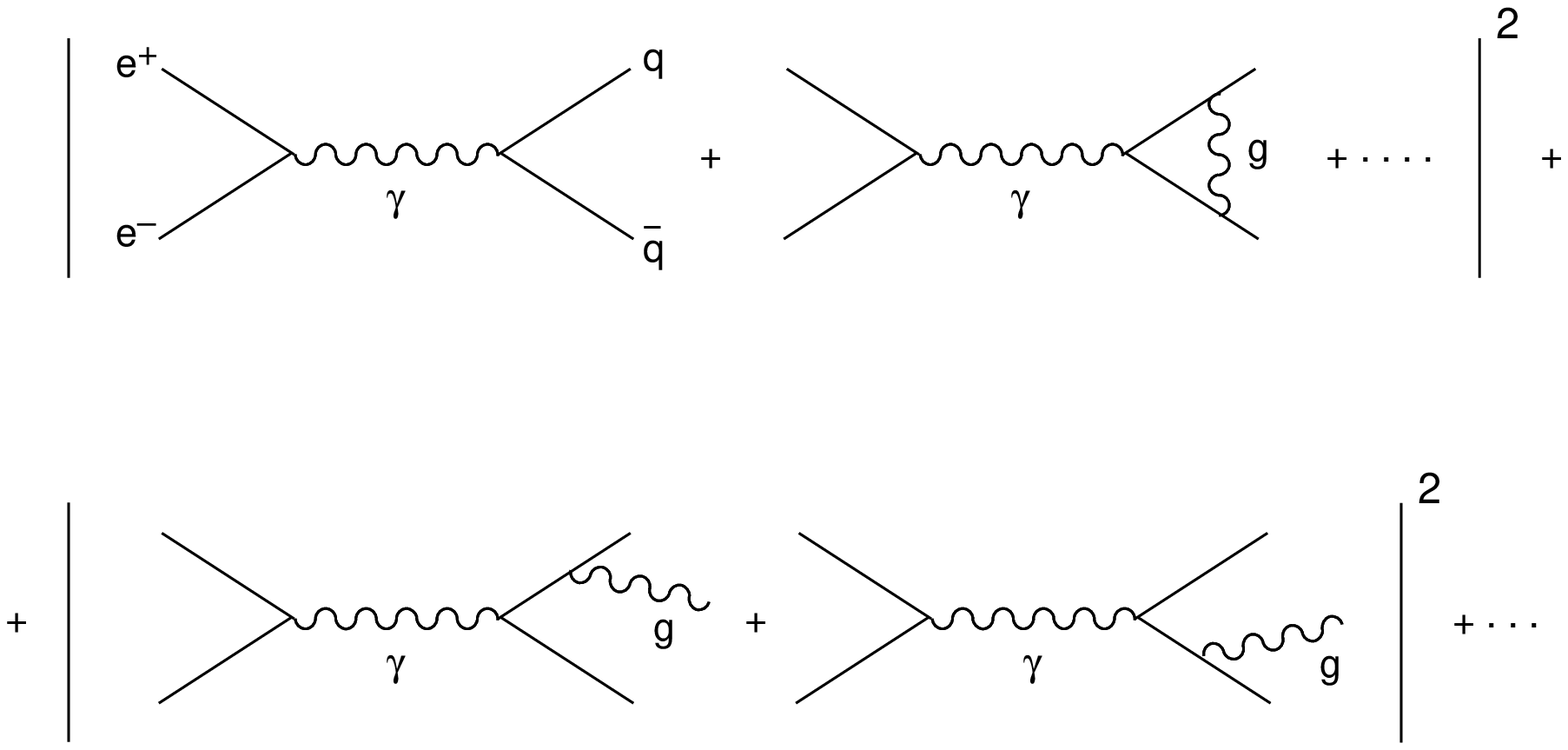}
\caption[]{}
\label{fig9}
\end{center}
\end{figure}

At 1-loop the diagrams relevant for the computation of R are shown in Fig.~9. There are virtual diagrams and real
diagrams with one additional gluon in the final state. Infrared divergences cancel  between the interference term of
the virtual diagrams and the absolute square of the real diagrams, according to the Bloch-Nordsieck theorem. Similarly
there are no mass singularities, in agreement with the Kinoshita-Lee-Nauenberg theorem, because the initial state is
purely leptonic and all degenerate states that can appear at the given order are included in the final state. Given that
$\gamma(\alpha_s)=0$ the RGE prediction is simply given, as we have already seen, by $F(t,\alpha_s)=F[0,\alpha_s(t)]$.
This means that if we do, for example, a 2-loop calculation, we must obtain a result of the form:
\beq
F(t,\alpha_s)~=~1~+~c_1\alpha_s(1-b\alpha_st)~+~c_2\alpha_s^2~+o(\alpha_s^3)\label{Fexp2}\\
\eeq  
In fact we see that this form, taking into account that from eq.(\ref{beQCD1}) we have:
\beq
\alpha_s(t)~\sim~\frac{\alpha_s}{1+b\alpha_s t}\sim\alpha_s(1~-~b\alpha_s t~+~....)\label{xx}\\
\eeq
can be rewritten as
\beq
F(t,\alpha_s)~=~1~+~c_1\alpha_s(t)~+~c_2\alpha_s^2(t)~+o(\alpha_s^3(t))~=~F[0,\alpha_s(t)]\label{firstFexp3}\\
\eeq  
The content of the RGE prediction is, at this order, that there are no $\alpha_st$ and $(\alpha_st)^2$ terms (the leading
log sequence must be absent) and the term of order $\alpha_s^2t$ has the coefficient that allows to reabsorb it in the
transformation of $\alpha_s$ into $\alpha_s(t)$.

At present the first 3 coefficients have been computed in the $\bar M \bar S$ scheme \cite{gor}. Clearly $c_1=1/\pi$ does not
depend on the definition of $\alpha_s$ but $c_2$ and $c_3$ do. The subleading coefficients also depend on the scale choice:
if instead of expanding in $\alpha_s(Q)$ we decide to choose $\alpha_s(Q/2)$ the coefficients $c_2$ and $c_3$ change. In
the  
$\bar M \bar S$ scheme, for $\gamma$-exchange and $n_f=5$, which are good approximations for $2m_b<<Q<<m_Z$, one has:
\beq
F[0,\alpha_s(t)]~=~1~+~\frac{\alpha_s(t)}{\pi}~+~1.409...(\frac{\alpha_s(t)}{\pi})^2~-~12.8...(\frac{\alpha_s(t)}{\pi})^3
~+...\label{Fexp3}\\
\eeq

Similar perturbative results at 3-loop accuracy also exist for $R_Z=\Gamma(Z\rightarrow hadrons)/\Gamma(Z\rightarrow
leptons)$, $R_{\tau}=\Gamma(\tau\rightarrow \nu_{\tau}+hadrons)/\Gamma(\tau\rightarrow \nu_{\tau}+ leptons)$, etc. We
will discuss these results later when we deal with measurements of $\alpha_s$.

The perturbative expansion in powers of $\alpha_s(t)$ takes into account all contributions that are suppressed by powers
of logarithms of the large scale $Q^2$ ("leading twist" terms). In addition there are corrections suppressed by powers of
the large scale $Q^2$ ("higher twist" terms).  The pattern of power corrections is controlled by the light-cone Operator
Product Expansion (OPE) which (schematically) leads to:
\beq
F~=~\rm{pert.}~+~r_2\frac{m^2}{Q^2}~+~r_4\frac{<0|Tr[\bf{F_{\mu\nu}}\bf{F^{\mu\nu}}]|0>}{Q^4}~+~...~+~
r_6\frac{<0|O_6|0>}{Q^6}~+~...\label{lce+e-}\\
\eeq
Here $m^2$ generically indicates mass corrections, notably from b quarks, for
example (t quark mass corrections only arise from loops, vanish in the limit $m_t\rightarrow \infty$ and are included in
the coefficients as those in eq.(\ref{Fexp3}) and the analogous ones for higher twist terms),
$\bf{F_{\mu\nu}}=\sum_AF_{\mu\nu}^At^A$, $O_6$ is typically a 4-fermion operator, etc. For each possible gauge invariant
operator the corresponding power of $Q^2$ is fixed by dimensions. 

We now consider the light-cone OPE in some more detail. $R_{e^+e^-}\sim \Pi(Q^2)$ where $\Pi(Q^2)$ is the scalar spectral
function related to the hadronic contribution to the imaginary part of the photon vacuum polarization $T_{\mu\nu}$:
\bea
T_{\mu\nu}~&=&~(-g_{\mu\nu}Q^2~+~q_{\mu}q_{\nu})\Pi(Q^2)~=~\int \exp{iqx} <0|J_{\mu}^\dagger (x)J_{\nu}(0)|0> dx~=~\nonumber\\
&&~=~\sum_n<0|J_{\mu}^\dagger (0)|n><n|J_{\nu}(0)|0>(2\pi)^4\delta^4(q-p_n)\nonumber\\
\label{Tmunu}\\
\nonumber\eea
For $Q^2\rightarrow \infty$ the $x^2\rightarrow 0$ region is dominant. To all orders in perturbation theory the OPE can be
proven. Schematically, dropping Lorentz indices, for simplicity, near $x^2\sim0$ we have:
\bea
J_{\mu}^\dagger (x)J_{\nu}(0)&=& I(x^2)~+~E(x^2)\sum_{n=0}^\infty c_n(x^2)x^{\mu_1}...x^{\mu_n}\cdot
O^n_{\mu_1...\mu_n}(0)~+~\nonumber\\
&&~+~\rm{less~sing.~terms}
\label{OPEJJ}\\
\nonumber\eea
Here $I(x^2)$, $E(x^2)$,..., $c_n(x^2)$ are c-number singular functions, $O^n$ is a string of local operators. $E(x^2)$ is
the singularity of free field theory, $I(x^2)$ and $c_n(x^2)$ contain powers of $\log{\mu x}$ in interaction. Some $O^n$
are already present in free field theory, other ones appear when interactions are switched on.
$\Pi(Q^2)$ is related to the Fourier transform. Less singular terms in $x^2$ lead to power suppressed terms in $1/Q^2$.   
The perturbative terms come from $I(x^2)$ which is the leading twist
term. The logarithmic scaling violations induced by the running coupling are the logs in $I(x^2)$.

\subsection{The Final State in $e^+e^-$ Annihilation}

Experiments on $e^+e^-$ annihilation at high energy provide a remarkable possibility of systematically testing the distinct
signatures predicted by QCD for the structure of the final state averaged over a large number of events. Typical of
asymptotic freedom is the hierarchy of configurations emerging as a consequence of the smallness of $\alpha_s(Q^2)$. When all
corrections of order $\alpha_s(Q^2)$ are neglected one recovers the naive parton model prediction for the final state:
almost collinear events with two back-to-back jets with limited transverse momentum and an angular distribution as
$(1+\cos^2{\theta})$ with respect to the beam axis (typical of spin 1/2 parton quarks: scalar quarks would lead to a
$\sin^2{\theta}$ distribution). At order $\alpha_s(Q^2)$ a tail of events is predicted to appear with large transverse
momentum $p_T\sim Q/2$ with respect to the thrust axis (the axis that maximizes the sum of the absolute values of the
longitudinal momenta of the final state particles). This small fraction of events with large $p_T$ mostly consists of
three-jet events with an almost planar topology. The skeleton of a three-jet event, at leading order in $\alpha_s(Q^2)$, is
formed by three hard partons $q\bar qg$, the third being a gluon emitted by a quark or antiquark line. The distribution of
three-jet events is given by:
\beq
\frac{1}{\sigma}\frac{d\sigma}{dx_1dx_2}~=~\frac{2\alpha_s}{3\pi}\frac{x_1^2+x_2^2}{(1-x_1)(1-x_2)}\label{3j}\\
\eeq
here $x_{1,2}$ refer to energy fractions of massless quarks: $x_i=2E_i/\sqrt{s}$ with $x_1+x_2+x_3=2$.
At order
$\alpha_s^2(Q^2)$ a hard perturbative non planar component starts to build up and a small fraction of four-jet events $q\bar
qgg$ or $q\bar q q\bar q$ appear, and so on.

For precise testing and for measuring $\alpha_s$ a quantitatively specified definition of jet counting must be introduced
which must be infrared safe (i.e. not altered by soft particle emission or collinear splittings of massless particles) in
order to be computable at parton level and as much as possible insensitive to the transformation of partons into hadrons. One
introduces a resolution parameter $y_{cut}$ and a suitable pair variable, for example:
\beq
y_{ij}~=~\frac{min(E_i^2,E_j^2)(1-\cos{\theta_{ij}})}{s}\label{yij}\\
\eeq
The particles i,j belong to different jets for $y_{ij}>y_{cut}$. Clearly the number of jets becomes a function of $y_{cut}$:
there are more jets for smaller $y_{cut}$. Measurements of $\alpha_s(Q^2)$ have been performed starting from jet
multiplicities, the largest error coming from the necessity of correcting for non-perturbative hadronisation effects.

\subsection{Deep Inelastic Scattering}

Deep Inelastic Scattering (DIS) processes
have played and still play a very important role for our understanding of QCD and of nucleon structure. This set of
processes actually provides us with a rich laboratory for theory and experiment. There are several structure functions that
can be studied, $F_i(x,Q^2)$, each a function of two variables. This is true separately for different beams and targets and
different polarizations. Depending on the charges of l and l' we can have neutral currents ($\gamma$,Z) or charged currents in
the l'-l channel (Fig.~6). In the past DIS processes were crucial for establishing quarks and gluons as partons and QCD as
the theory of strong interactions. At present DIS is very important for quantitative studies and tests of QCD. The theory
of scaling violations for totally inclusive DIS structure functions, based on operator expansions and renormalization group
techniques, is crystal clear and the predicted $Q^2$ dependence can be tested at each value of x. The measurement of quark and
gluon densities in the nucleon, as functions of x at some reference value of $Q^2$, which is an essential starting point for
the calculation of all relevant hadronic hard processes, is performed in DIS processes. At the same time one measures
$\alpha_s(Q^2)$ and the DIS values can be compared with those obtained from other processes. At all times new theoretical
challenges arise from the study of DIS processes. Recent examples are the so-called "spin crisis" in polarized DIS and the
behaviour of singlet structure functions at small x as revealed by HERA data. In the following we will review the past
successes and the present open problems in the physics of DIS.

The cross-section $\sigma\sim L^{\mu \nu}W_{\mu \nu}$ is given in terms of the product of a leptonic ($L^{\mu \nu}$) and a
hadronic ($W_{\mu
\nu}$) tensor. While $L^{\mu \nu}$ is simple and easily obtained from the lowest order electroweak (e-w) vertex plus QED
radiative corrections, the complicated strong interaction dynamics is contained in $W_{\mu \nu}$. The latter is proportional
to the Fourier transform of the forward matrix element between the nucleon target states of the product of two e-w
currents:
\beq W_{\mu \nu}~=~\int{~dx~\exp{iqx}~<p|J^{\dagger}_{\mu}(x)J_{\nu}(0)|p>}\label{FTx}
\eeq Structure functions are defined starting from the general form of $W_{\mu \nu}$ given Lorentz invariance and current
conservation. For example, for e-w currents between unpolarized nucleons we have:
\bea
W_{\mu \nu}~&=&~(-g_{\mu \nu}~+~\frac{q_{\mu}q_{\nu}}{q^2})~W_1(\nu,Q^2)~+~(p_{\mu}~-~\frac{m
\nu}{q^2}q_{\mu})(p_{\nu}~-~\frac{m
\nu}{q^2}q_{\nu})~\frac{W_2(\nu,Q^2)}{m^2}~-~\nonumber\\
&&~-~\frac{i}{2m^2}\epsilon_{\mu \nu \lambda
\rho}p^{\lambda}q^{\rho}~W_3(\nu,Q^2)\nonumber\\
\label{sf}
\nonumber\eea
$W_3$ is absent for pure vector currents. In the limit $Q^2>>m^2$, x fixed, the structure functions obey approximate Bjorken
scaling which in reality is broken by logarithmic corrections that can be computed in QCD:
\bea mW_1(\nu,Q^2)\rightarrow F_1(x)\nonumber \\
\nu W_{2,3}(\nu,Q^2)\rightarrow F_{2,3}(x) \label{Bj}
\eea 
The $\gamma-N$ cross-section is given by ($W_i~=~W_i(Q^2,\nu)$):
\beq
\frac{d\sigma^{\gamma}}{dQ^2d\nu}~=~\frac{4\pi\alpha^2E'}{Q^4E}\cdot
[2\sin^2{\theta/2}W_1~+~\cos^2{\theta/2}W_2]\label{gN}\\
\eeq
while for the $\nu-N$ or $\bar{\nu}-N$ cross-section one has:
\beq
\frac{d\sigma^{\nu,\bar{\nu}}}{dQ^2d\nu}~=~\frac{G_F^2E'}{2\pi E}(\frac{m_W^2}{Q^2+m_W^2})^2\cdot
[2\sin^2{\theta/2}W_1~+~\cos^2{\theta/2}W_2\pm\frac{E+E'}{m}\sin^2{\theta/2}W_3]\label{nuN}\\
\eeq 
($W_i$ for photons, $\nu$ and $\bar{\nu}$ are all different, as we shall see in a moment). 

In the scaling limit the longitudinal and transverse cross sections are given by:
\bea
\sigma_L~&=&~\frac{1}{s}[\frac{F_2(x)}{2x}~-~F_1(x)]\nonumber \\
&&\sigma_{RH,LH}\sim \frac{1}{s}[F_1(x)~\pm~F_3(x)]\nonumber \\
&&\sigma_T~=~\sigma_{RH}~+~\sigma_{LH} \label{sig}
\eea where L, RH, LH refer to the helicity 0, 1, -1, respectively, of the exchanged gauge vector boson.

In the '60's the demise of hadrons from the  status of fundamental particles to that of bound states of constituent quarks was
the breakthrough that made possible the construction of a renormalisable field theory for strong interactions. The presence of
an unlimited number of hadrons species, many of them with large spin values, presented an obvious dead-end for a manageable field
theory. The evidence for constituent quarks emerged clearly from the systematics of hadron spectroscopy. The complications of
the hadron spectrum could be explained in terms of the quantum numbers of spin 1/2, fractionally charged, u, d and s quarks.
The notion of colour was introduced to reconcile the observed spectrum with Fermi statistics. But confinement that forbids 
the
observation of free quarks was a clear obstacle towards the acceptance of quarks as real constituents and not just as
fictitious entities describing some mathematical pattern (a doubt expressed even by Gell-Mann at the time). The early
measurements at SLAC of DIS dissipated all doubts: the observation of Bjorken scaling and the success of the naive (not so
much after all) parton model of Feynman imposed quarks as the basic fields for describing the nucleon structure (parton
quarks). 

In the language of Bjorken and Feynman the virtual $\gamma$ (or, in general, any gauge boson) sees the quark partons
inside the nucleon target as quasi-free, because the (Lorentz dilated) QCD interaction time is much longer than
$\tau_{\gamma}\sim 1/Q$. Since the virtual photon 4-momentum is spacelike, we can go to a Lorentz frame where $E_{\gamma}=0$
(Breit frame). In this frame $q=(E_{\gamma}=0;0,0,Q)$ and the nucleon momentum, neglecting the mass $m<<Q$, is
$p=(Q/2x;0,0,-Q/2x)$ (note that this correctly gives $x=Q^2/2(p\cdot q)$). Consider (Fig.~10) the interaction of the photon
with a quark carrying a fraction y of the nucleon 4-momentum: $p_q=yp$ (we are neglecting the transverse components of $p_q$
which are of order $m$). The incoming parton with $p_q=yp$ absorbs the photon and the final parton has 4-momentum
$p'_q$. Since in the Breit frame the photon carries no energy but only a longitudinal momentum $Q$, the photon can only be 
absorbed by those partons with $y=x$: then the longitudinal component of $p_q=yp$ is $-yQ/2x=-Q/2$ and can be flipped into
$+Q/2$ by the photon. As a result, the photon longitudinal momentum $+Q$ disappears, the parton quark momentum changes of
sign from
$-Q/2$ into $+Q/2$ and the energy is not changed. So the structure functions are proportional to the density of partons
with fraction x of the nucleon momentum, weighted with the squared charge. Also, recall that the helicity of a massless quark
is conserved in a vector (or axial vector) interaction. So when the momentum is reversed also the spin must flip. Since the
process is collinear there is no orbital contribution and only a photon with helicity
$\pm 1$ (transverse photon) can be absorbed. If partons were spin zero only longitudinal photons would instead contribute. 

\begin{figure}[h]
\begin{center}
\includegraphics[width=5cm]{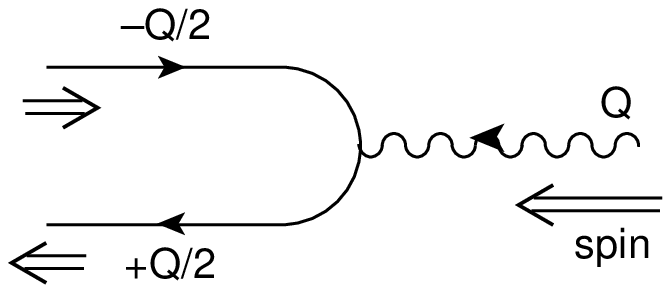}
\caption[]{}
\label{fig10}
\end{center}
\end{figure}

Using these results, which are maintained in QCD at leading order, the quantum numbers of the quarks were confirmed by
early experiments. The observation that
$R~=~\sigma_L/\sigma_T\rightarrow 0$ implies that the charged partons have spin 1/2. The quark charges were derived from the
data on the electron and neutrino structure functions:
\bea
F_{ep}~=~4/9u(x)~+~1/9d(x)~+~.....;~~~~~~F_{en}~=~4/9d(x)~+~1/9u(x)~+~....\nonumber\\
F_{\nu p}~=~F_{\bar{\nu}n}~=~2d(x)~+~.....;~~~~~~F_{\nu n}~=~F_{\bar{\nu}p}~=~2u(x)~+~.....\label{charges}
\eea
where $F\sim 2F_1\sim F_2/x$ and $u(x)$, $d(x)$ are the parton number densities in the proton (with fraction x of the proton
longitudinal momentum), which, in the scaling limit, do not depend on $Q^2$. The normalisation of the structure functions
and the parton densities are such that the charge relations hold:
\beq
\int_0^1[u(x)-\bar u(x)]dx=2,~~~\int_0^1[d(x)-\bar d(x)]dx=1,~~~\int_0^1[s(x)-\bar s(x)]dx=0\label{cha}\\
\eeq
Also it was proven by experiment that at values of
$Q^2$ of a few $GeV^2$, in the scaling region, about half of the nucleon momentum, given by the momentum sum rule:
\beq
\int_0^1[\sum_i(q_i(x)+\bar{q}_i(x))~+~g(x)]xdx~=~1\label{momsr}\\
\eeq
is carried by neutral partons (gluons).

In QCD there are calculable log scaling violations induced by $\alpha_s(t)$. The parton rules just introduced can be
summarised in the formula:
\beq
F(x,t)~=~\int_x^1dy\frac{q_0(y)}{y}\sigma_{point}(\frac{x}{y},\alpha_s(t))~+~o(\frac{1}{Q^2})\label{conv1}\\
\eeq
Before QCD corrections $\sigma_{point}=e^2\delta(x/y-1)$ and $F=e^2q_0(x)$ (here we denote by $e$ the charge of the quark in
units of the positron charge, i.e. $e=2/3$ for the $u$ quark). QCD modifies
$\sigma_{point}$ at order
$\alpha_s$ via the diagrams of Fig.~11. Note that the integral is from x to 1, because the energy can only be lost by
radiation before interacting with the photon (which eventually wants to find a fraction $x$, as we have explained). From a
direct computation of the diagrams one obtains a result of the following form:
\beq
\sigma_{point}(z,\alpha_s(t))~\simeq ~e^2[\delta (z-1)~+~\frac{\alpha_s}{2\pi}(t\cdot P(z)~+~f(z))]\label{sigalf}\\
\eeq
\begin{figure}[h]
\begin{center}
\includegraphics[width=8cm]{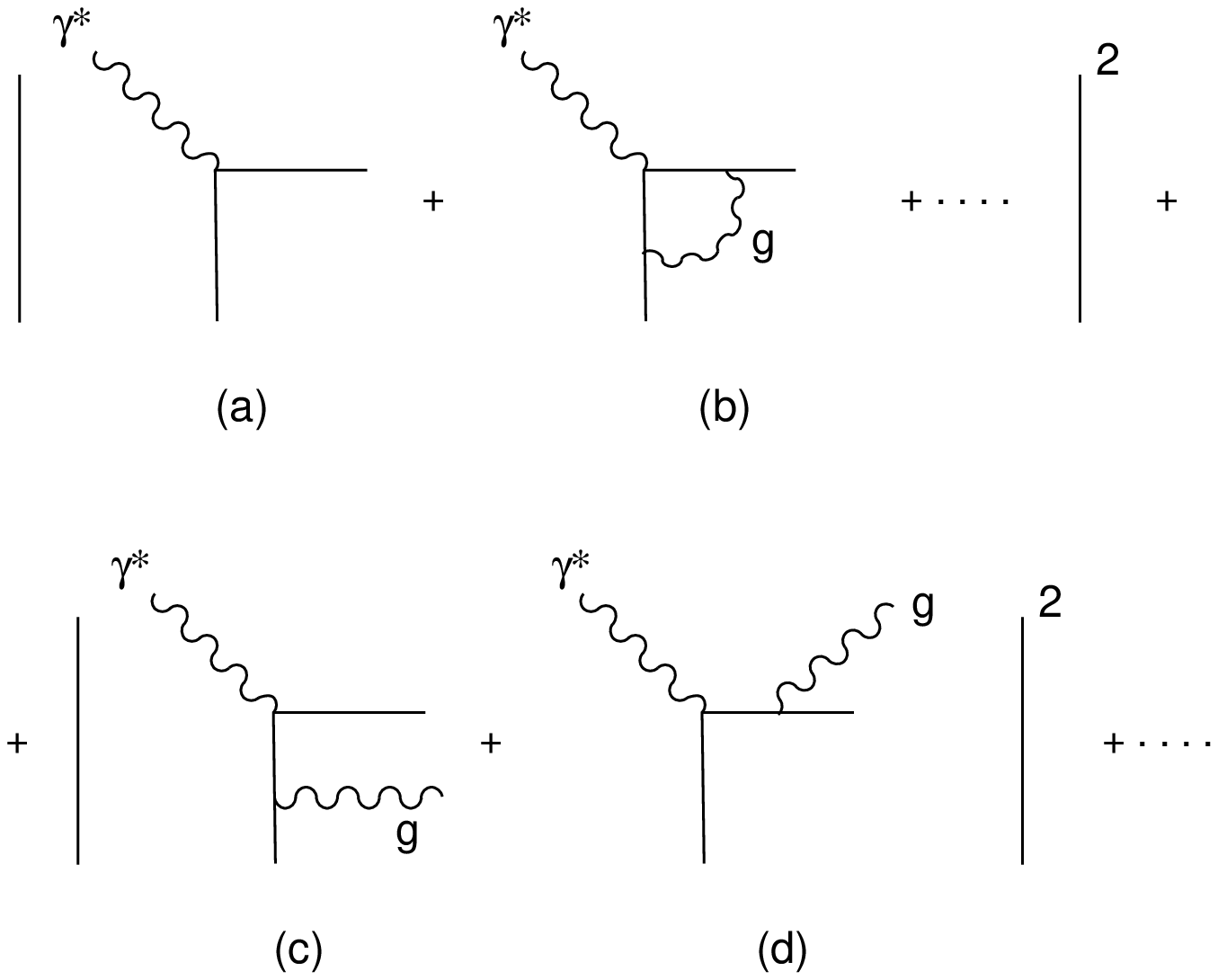}
\caption[]{}
\label{fig11}
\end{center}
\end{figure}

For $y>x$ the correction arises from diagrams with real gluon emission. Only the sum of the two real diagrams in Fig.~11 is
gauge invariant, so that the contribution of one given diagram is gauge dependent. There is a special form of axial gauge,
called physical gauge, where, among real diagrams, the diagram of Fig.~11(c) gives the whole $t$-proportional term. It is
obviously not essential to go to this gauge, but this diagram has a direct physical interpretation: a quark in the proton
has a fraction
$y>x$ of the parent 4-momentum; it then radiates a gluon and looses energy down to a fraction $x$ before interacting with
the photon. The log arises from the virtual quark propagator, according to the discussion of collinear mass singularities in
eq.(\ref{prop}). In fact in the massless limit one has:
\bea
\rm{propagator}~&=&~\frac{1}{r^2}~=~\frac{1}{(k-h)^2}~=~\frac{-1}{2E_kE_h}\cdot\frac{1}{1-\cos{\theta}}\nonumber\\
&&~=~\frac{-1}{4E_kE_h}\cdot\frac{1}{\sin^2{\theta/2}}~\propto \frac{-1}{p_T^2}
\label{prop1}
\eea 
where $p_T$ is the transverse momentum of the virtual quark. So the square of the propagator goes like $1/p_T^4$. But there
is a $p_T^2$ factor in the numerator, because in the collinear limit, when $\theta=0$ and the initial and final quarks and
the emitted gluon are all aligned, the quark helicity cannot flip (vector interaction) and the real gluon cannot have but
$\pm 1$ helicity. So the cross-section behaves as:
\beq
\sigma~\sim ~\int^{Q^2}\frac{1}{p_T^2}dp_T^2~\sim ~\log{Q^2}\label{log}\\
\eeq
Actually the log should be read as $\log{Q^2/m^2}$ because in the massless limit a genuine mass singularity appears. In fact
the mass singularity connected with the initial quark line is not cancelled because we do not have the sum of all degenerate
initial states, but only a single quark. But in correspondence to the initial quark we have the (bare) quark density
$q_0(y)$ that appear in the convolution integral. This is a non perturbative quantity that is determined by the nucleon wave
function. So we can factorize the mass singularity in a redefinition of the quark density: we replace $q_0(y)\rightarrow
q(y,t)~=~q_0(y)~+~\Delta q(y,t)$ with:
\beq
\Delta q(y,t)~=~\frac{\alpha_s}{2\pi}t\int_x^1dy\frac{q_0(y)}{y}\cdot P(\frac{x}{y})\label{deltaq}\\
\eeq
Here the factor of $t$ is a bit symbolic: it stands for $\log{Q^2/km^2}$ and what we exactly put below $Q^2$ depends on the
definition of the renormalised quark density, that also determines the exact form of the finite term $f(z)$ in
eq.(\ref{sigalf}).

The effective parton density $q(y,t)$ that we have defined is now scale dependent. In terms of this scale dependent density
we have the following relations, where we have also replaced the fixed coupling with the running coupling according to the 
prescription derived from the RGE:
\bea
F(x,t)~=~\int_x^1dy\frac{q(y,t)}{y}e^2[\delta
(\frac{x}{y}-1)~+~\frac{\alpha_s(t)}{2\pi}f(\frac{x}{y}))]~=~e^2q(x,t)~+~o(\alpha_s(t))\nonumber\\
\frac{d}{dt}q(x,t)~=~\frac{\alpha_s(t)}{2\pi}\int_x^1dy\frac{q(y,t)}{y}\cdot P(\frac{x}{y})~+~o(\alpha_s(t)^2)\label{APNS}\\
\nonumber\eea
We see that in lowest order we reproduce the naive parton model formulae for the structure functions in terms of effective
parton densities that are scale dependent. The evolution equations for the parton densities are written down in terms of
kernels (the "splitting functions") that can be expanded in powers of the running coupling. At leading order, we can
interpret the evolution equation by saying that the variation of the quark density at $x$ is given by the convolution of the
quark density at $y$ times the probability of emitting a gluon with fraction $x/y$ of the quark momentum.

It is interesting that the integro-differential QCD evolution equation for densities can be transformed into an ordinary
differential equation for Mellin moments. The moment $f_n$ of a density $f(x)$ is defined as:
\beq
f_n~=~\int_0^1dxx^{n-1}f(x)\label{Mel}\\
\eeq
By taking moments of both sides of the second of eqs.(\ref{APNS}) one finds, with a simple interchange of the integration
order, the simpler equation for the $nth$ moment:
\beq
\frac{d}{dt}q_n(t)~=~\frac{\alpha_s(t)}{2\pi}\cdot P_n \cdot q_n(t)\label{momev}\\
\eeq
To solve this equation we observe that:
\beq
\log{\frac{q_n(t)}{q_n(0)}}~=~\frac{P_n}{2\pi}\int_0^t\alpha_s(t)dt~=~\frac{P_n}{2\pi}\int_{\alpha_s}^{\alpha_s(t)}
\frac{d\alpha'}{-b\alpha'}\label{nn}\\
\eeq
where we used eq.(\ref{runt}) to change the integration variable from $dt$ to $d\alpha(t)$ (denoted as $d\alpha'$) and
$\beta(\alpha)\simeq -b\alpha^2+...$. Finally the solution is:
\beq
q_n(t)~=~[\frac{\alpha_s}{\alpha_s(t)}]^{\frac{P_n}{2\pi b}}\cdot q_n(0)\label{solmom}\\
\eeq
 
The connection of these results with the RGE general formalism occurs via the light cone OPE (recall eq.(\ref{FTx}) for
$W_{\mu\nu}$ and eq.(\ref{OPEJJ}) for the OPE of two currents). In the case of DIS the c-number term $I(x^2)$ does not
contribute, because we are interested in the connected part $<p|...|p>-<0|...|0>$. The relevant terms are:
\beq
J_{\mu}^\dagger (x)J_{\nu}(0)~=~E(x^2)\sum_{n=0}^\infty c_n(x^2)x^{\mu_1}...x^{\mu_n}\cdot
O^n_{\mu_1...\mu_n}(0)~+~
\rm{less~sing.~terms}
\label{OPE}
\eeq
A formally intricate but conceptually simple argument (Ref.\cite{Book}, page~28) based on the analiticity properties of the forward virtual Compton
amplitude shows that the Mellin moments $M_n$ of structure functions are related to the individual terms in the OPE,
precisely to the Fourier transform $c_n(Q^2)$ (we will write it as $c_n(t,\alpha)$)  of the coefficient $c_n(x^2)$ times a
reduced matrix element
$h_n$ from the operators
$O^n$:
$<p|O^n_{\mu_1...\mu_n}(0)|p>=h_n p_{\mu_1}...p_{\mu_n}$:
\beq
c_n<p|O^n|p>\rightarrow M_n=\int_0^1dxx^{n-1}F(x)\label{MomOPE}\\
\eeq
Since the matrix element of the products of currents satisfy the RGE so do the moments $M_n$. Hence the general form of the
$Q^2$ dependence is given by the RGE solution (see eq.(\ref{Fsol2})):
\beq
M_n(t,\alpha)~=~c_n[0,\alpha(t)]\exp{\int_{\alpha}^{\alpha(t)}\frac{\gamma_n(\alpha')}{\beta(\alpha')}d\alpha'}\cdot
h_n(\alpha)\label{Msol}\\
\eeq
In lowest order, identifying in the simplest case $M_n$ with $q_n$, we have:
\beq
\gamma_n(\alpha)~=~\frac{P_n}{2\pi} \alpha~+~...,~~~~~~~~~\beta(\alpha)~=~-b\alpha^2~+~...\label{Msol1}\\
\eeq
and  
\beq
q_n(t)=q_n(0)\exp{\int_{\alpha}^{\alpha(t)}\frac{\gamma_n(\alpha')}{\beta(\alpha')}d\alpha'}~=~
[\frac{\alpha_s}{\alpha_s(t)}]^{\frac{P_n}{2\pi b}}\cdot q_n(0)\label{solmom2}\\
\eeq
which exactly coincides with eq.(\ref{solmom}).

Up to this point we have implicitely restricted our attention to non-singlet (under the flavour group) structure
functions. The $Q^2$ evolution equations become non diagonal as soon as we take into account the presence of gluons in the
target. In fact the quark which is seen by the photon can be generated by a gluon in the target (Fig.~12). 

\begin{figure}[h]
\begin{center}
\includegraphics[width=5cm]{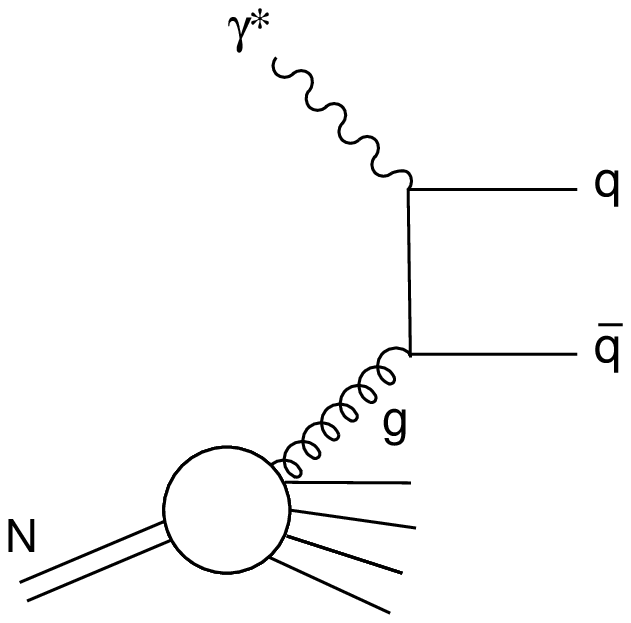}
\caption[]{}
\label{fig12}
\end{center}
\end{figure}

The quark evolution
equation becomes:
\beq
\frac{d}{dt}q_i(x,t)~=~\frac{\alpha_s(t)}{2\pi}[q_i\bigotimes P_{qq}]~+~\frac{\alpha_s(t)}{2\pi}[g\bigotimes
P_{qg}]\label{APqS}\\
\eeq
where we introduced the shorthand notation:
\beq
[q\bigotimes P]~=~[P\bigotimes q]~=~\int_x^1dy\frac{q(y,t)}{y}\cdot P(\frac{x}{y})\label{convdef}\\
\eeq
(it is easy to check that the convolution, like an ordinary product, is commutative). At leading order, the interpretation of
eq.(\ref{APqS}) is simply that the variation of the quark density is due to the convolution of the quark density at a higher
energy times the probability of finding a quark in a quark (with the right energy fraction) plus the gluon density  at a
higher energy times the probability of finding a quark (of the given flavour i) in a gluon. The evolution equation for the
gluon density, needed to close the system, can be obtained by suitably extending the same line of reasoning to a gedanken
probe sensitive to colour charges, for example a virtual gluon. The resulting equation is of the form:
\beq
\frac{d}{dt}g(x,t)~=~\frac{\alpha_s(t)}{2\pi}[\sum_i (q_i+\bar q_i)\bigotimes P_{gq}]~+~\frac{\alpha_s(t)}{2\pi}[g\bigotimes
P_{gg}]\label{APgS}\\
\eeq

The explicit form of the splitting functions in lowest order can be explicitly derived from the QCD vertices. They are a
property of the theory and do not depend on the particular process the parton density is taking part into. The results are:
\bea
P_{qq}~=~\frac{4}{3}[\frac{1+x^2}{(1-x)_+}~+~\frac{3}{2}\delta (1-x)]~+~o(\alpha_s)\nonumber\\
P_{gq}~=~\frac{4}{3}\frac{1+(1-x)^2}{x}~+~o(\alpha_s)\nonumber\\
P_{qg}~=~\frac{1}{2}[x^2+(1-x)^2]~+~o(\alpha_s)\nonumber\\
P_{gg}~=~6[\frac{x}{(1-x)_+}~+~\frac{1-x}{x}~+~x(1-x)]~+~\frac{33-2n_f}{6}\delta (1-x)~+~o(\alpha_s)
\label{Splfs}
\eea
The "+" distribution is defined as, for a generic non singular weight function $f(x)$:
\beq
\int_0^1\frac{f(x)}{(1-x)_+}dx~=~\int_0^1\frac{f(x)-f(1)}{1-x}dx\label{plus}\\
\eeq
The $\delta(1-x)$ terms arise from the virtual corrections to the tree diagram. Their coefficient can be simply obtained by
imposing the validity of charge and momentum sum rules. In fact, from the request that the charge sum rules in
eq.(\ref{cha}) are not affected by the $Q^2$ dependence one derives that
\beq
\int_0^1P_{qq}(x)dx~=~0\label{Pqqsr}\\
\eeq
which can be used to fix the coefficient of the $\delta(1-x)$ terms of $P_{qq}$. Similarly, by taking the t-derivative of the
momentum sum rule in eq.(\ref{momsr}) and imposing its vanishing for generic $q_i$ and $g$, one obtains:
\beq
\int_0^1[P_{qq}(x)~+~P_{gq}(x)]xdx~=~0,~~~~~~\int_0^1[2n_fP_{qg}(x)~+~P_{gg}(x)]xdx~=~0,\label{Pmomsr}\\
\eeq

At higher orders the evolution equations are easily generalised but the calculation of the splitting functions rapidly
becomes very complicated. The splitting functions are completely known at NLO accuracy: $\alpha_s P~\sim~\alpha_s
P_1~+~\alpha_s^2 P_2~+...$. More recently the NNLO results
$P_3$ have been derived in analytic form for the first few moments \cite{ver}. The full NNLO calculation is in progress and
could be finished soon. 

The scaling violations are clearly observed by experiment and their pattern is very well reproduced by QCD fits at
NLO. Examples are seen in Figs.~13(a-d) \cite{rob}. These fits provide an impressive confirmation of a quantitative QCD
prediction, a measurement of $q_i(x,Q_0^2)$ and $g(x,Q_0^2)$ at some reference value $Q_0^2$ of $Q^2$ and a precise
measurement of
$\alpha_s(m_Z^2)$. At small $x$ and large but fixed $Q^2$ when terms of order $(\alpha_s \log{1/x})^n$ become of order 1 and
cannot be neglected, the validity of the NLO or NNLO approximations should breakdown. However, the small $x$ data collected
by HERA can be fitted reasonably even at the smallest measured values of $x$ by the NLO QCD evolution equations, so that
there is no neat evidence in the data for departures. But the extracted gluon density (which is dominant at small $x$) and
the fitted value of $\alpha_s(m_Z^2)$ could be biassed by this effect if data at too small values of $x$ are included.

\begin{figure}[htbp]
\begin{center}
\includegraphics[width=15cm]{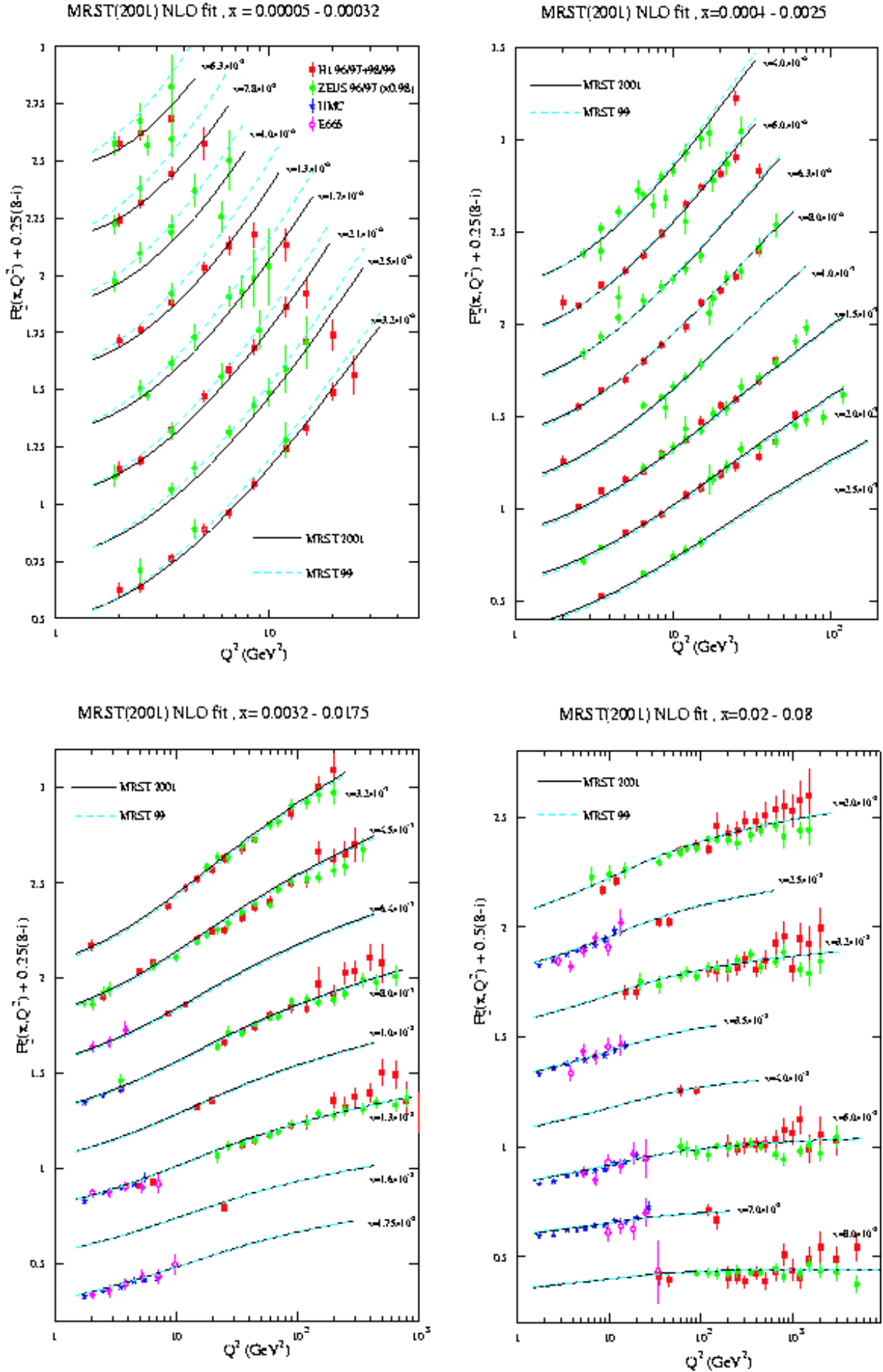}
\caption[]{A recent NLO fit of scaling violations from Ref.~\cite{rob}, for different $x$ ranges, as functions of $Q^2$}
\label{fig13}
\end{center}
\end{figure}
\subsection{Factorisation and the QCD Improved Parton Model}

The parton densities defined and measured in DIS are used to compute hard processes initiated by hadronic collisions via the
Factorisation Theorem (FT). Suppose you have a hadronic process of the form $h_1+h_2 \rightarrow X+all$ where $X$ is some
triggering particle or pair of particles which specify the large scale $Q^2$ relevant for the process, in general somewhat
but not much smaller than s, the total c.o.m. squared mass. For example, in
$pp$ or
$p\bar p$ collisions, $X$ can be a $W^±$ or a Z or a virtual photon with large $Q^2$, or a jet at large transverse momentum
$p_T$, or a pair of heavy quark-antiquark of mass M. By "all" we mean a totally inclusive collection of gluons and light
quark pairs. The FT implies that for the cross-section or some other sufficiently inclusive distribution we can write the
expression:
\beq
\sigma(s,\tau)~=~\sum_{AB}\int dx_1dx_2 p_{1A}(x_1,Q^2)p_{2B}(x_2,Q^2)\sigma_{AB}(x_1x_2s,\tau)\label{FT}\\
\eeq
Here $\tau=Q^2/s$ is a scaling variable, $p_{iC}$ are generic parton-C densities inside the hadron $h_i$, $\sigma_{AB}$
is the partonic cross-section for parton-A + parton-B$\rightarrow X + all'$. This result is based on the fact that the mass
singularities that are associated with the initial legs are of universal nature, so that one can reproduce the same modified
parton densities, by absorbing these singularities into the bare parton densities, as in deep inelastic scattering. Once the
parton densities and $\alpha_s$ are known from other measurements the prediction of the rate for a given hard process is
obtained with no free parameters. The NLO calculation of the reduced partonic cross-section is needed in order to
correctly specify the scale and in general the definition of the parton densities and of the running coupling in the leading
term. The residual scale and scheme dependence is often the most important source of theoretical error. In the following we
consider a few examples.

A comparison of data and predictions on the production of jets at large $\sqrt s$ and $p_T$ in $pp$ or
$p\bar p$ collisions is shown in Fig.~14 \cite{pdg}.

\begin{figure}[htbp]
\begin{center}
\includegraphics[width=8cm]{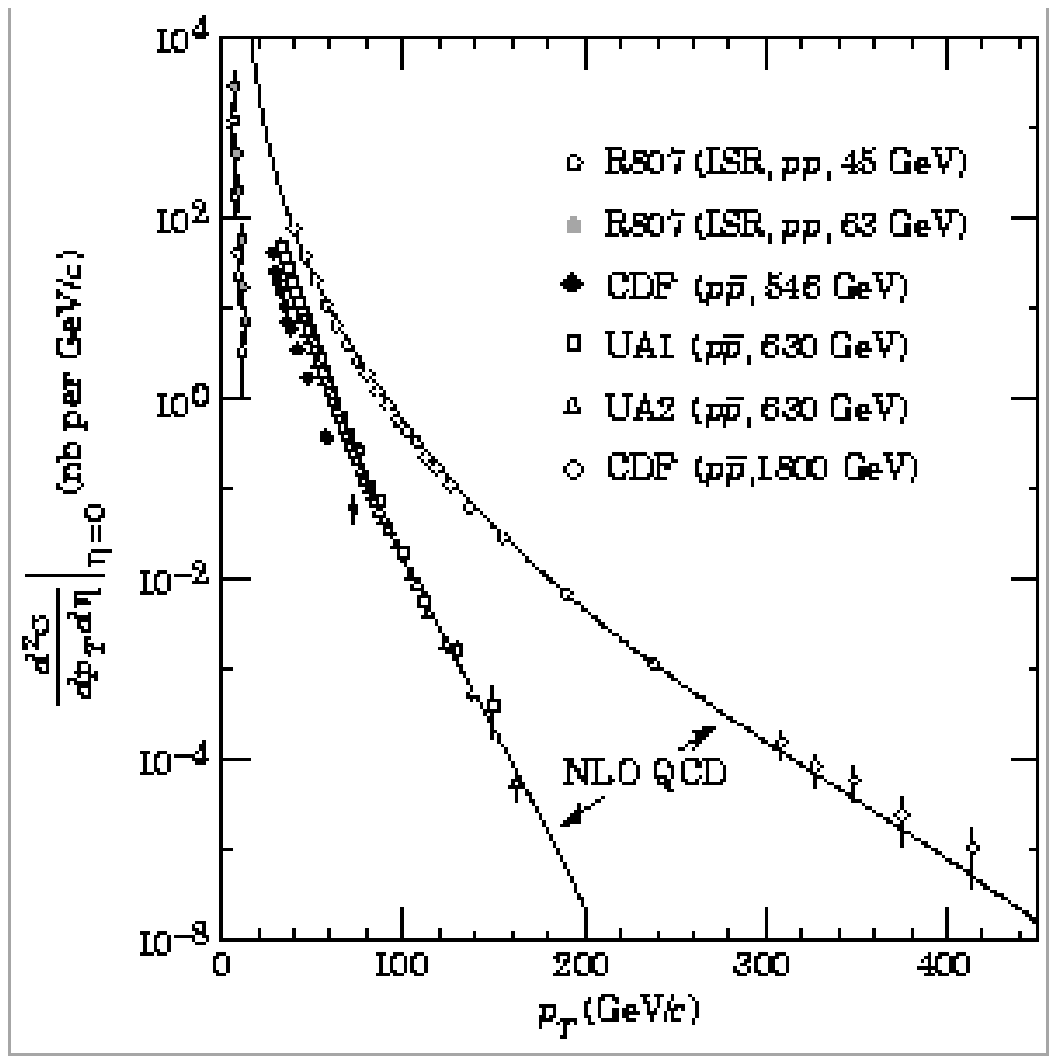}
\caption[]{Jet production cross-section at $pp$ or $p\bar p$ colliders, as function of $p_T$}
\label{fig14}
\end{center}
\end{figure}

This is a particularly significant test because the rates at different
c.o.m. energies and, for each energy, at different values of $p_T$ span over many orders of magnitude. Also the
corresponding values of $\sqrt s$ and $p_T$ are large enough to be well inside the perturbative region. The overall
agreement of the data from ISR, UA1,2 and CDF and D0 is spectacular. Only at very large $p_T$ there might be some problem
according to CDF and, to a lesser extent, to D0. A harder gluon at large x and inclusion of systematic errors can
alleviate or eliminate the problem. This issue will be clarified in the near future by the RunII results at the Tevatron.
Similar results also hold for the production of photons at large $p_T$. The collider data, shown in Fig.~15 \cite{pdg}, are
in fair agreement with the theoretical predictions. A less clear situation is found with fixed target data. Here, first of
all, the experimental results show some internal discrepancies. Also, the $p_T$ accessible values being smaller, the
theoretical uncertainties are larger. But it is true that the agreement is poor, so that the necessity of an "intrinsic"
transverse momentum of partons inside the hadron of over 1 GeV has been claimed, which I do not think one is free to
introduce.
\begin{figure}[h]
\begin{center}
\includegraphics[width=8cm]{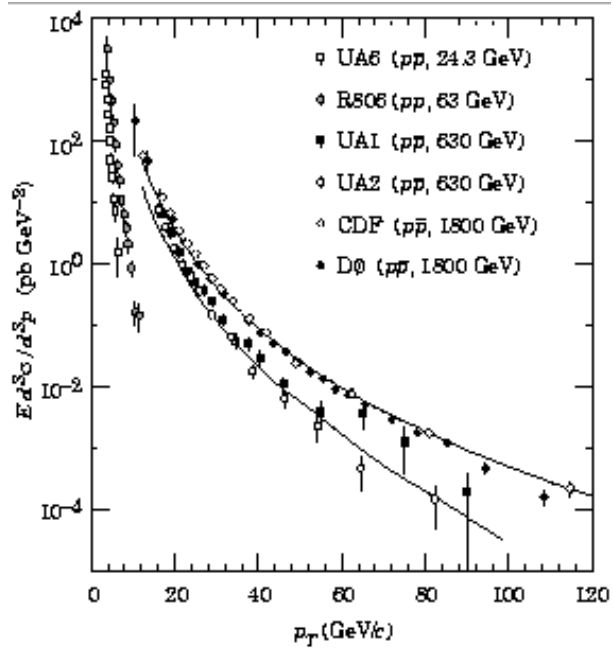}
\caption[]{Single photon production in $p\bar p$ colliders as function of $p_T$}
\label{fig15}
\end{center}
\end{figure}

For heavy quark production at colliders the agreement is very good for top production at the Tevatron (Fig.~16). To present a
rare example of a puzzle, the bottom production at the Tevatron remains problematic. The total rate and the $p_T$
distribution of b quarks observed at CDF is in excess of the prediction, up to the largest measured values of $p_T$ (fig.17).
It is true that this is a complicated case, with different scales being
present at the same time: $\sqrt{s}$, $p_T$, $m_b$.  Also some non-perturbative ingredients, like fragmentation functions, are present in the calculation and could be in part reponsible for the effect.  But no
clear explanation of this phenomenon is available. Probably it occurs through an unfavourable conspiracy of several small
effects.

\begin{figure}[h]
\begin{center}
\includegraphics[width=8cm]{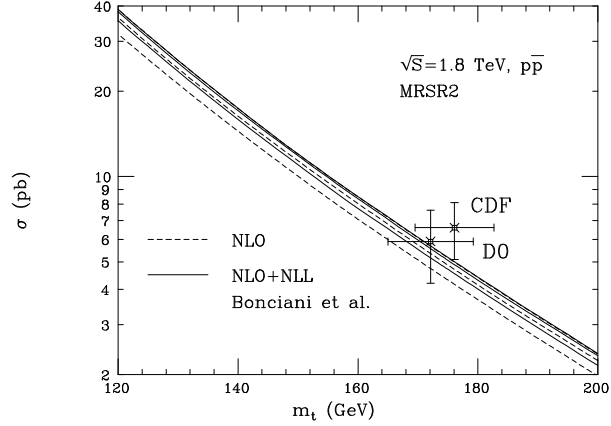}
\caption[]{The $t$ production cross-section at the Tevatron $p\bar p$ collider}
\label{fig16}
\end{center}
\end{figure}

\begin{figure}[h]
\begin{center}
\includegraphics[width=8cm]{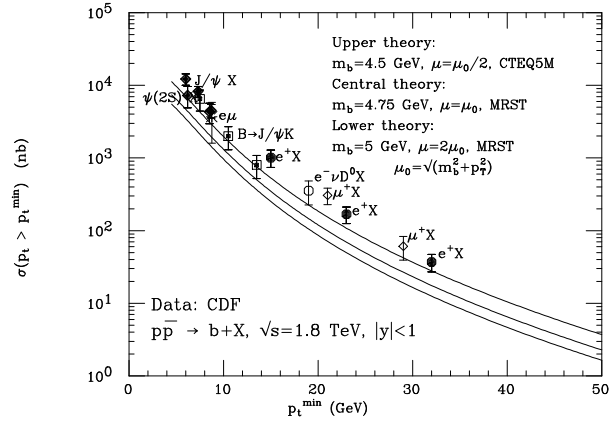}
\caption[]{The $b$ production $p_T$ distribution at the Tevatron $p \bar p$ collider (CDF collaboration).  The cross-section for $p_T > p^{min}_T$ is plotted as function of $p^{min}_T$.}
\label{fig17}
\end{center}
\end{figure}

Drell-Yan processes, including lepton pair production via virtual $\gamma$, W or Z exchange, offer a good opportunity to
test QCD. The process is quadratic in parton densities, and the final state is totally inclusive, while the large scale is
specified and measured by the invariant mass squared $Q^2$ of the lepton pair which itself is not strongly interacting. The
QCD improved parton model leads directly to a prediction for the total rate as a function of $Q^2$. The value of the LO
cross-section is inversely proportional to the number of colours $N_C$ because a quark of given colour can only annihilate
with an antiquark of the same colour to produce a colourless lepton pair. The order $\alpha_s(Q^2)$ corrections to the
total rate were computed long ago and found to be particularly large, when the quark densities are defined from
the structure function $F_2$ measured in DIS at $q^2=-Q^2$. The ratio $\sigma_{corr}/\sigma_{LO}$ of the corrected and the
Born cross-sections, was called K-factor (by me), because it is almost a constant in rapidity. More recently also the NLO full
calculation of the K-factor was completed, a very remarkable calculation \cite{nee}. The QCD predictions can be best tested
for W and Z production at CERN and Tevatron energies. $Q\sim m_{W,Z}$ is large enough to make the prediction reliable (a not
too large K-factor) and the ratio $\sqrt{\tau}=Q/\sqrt{s}$ is not too small. Recall that in lowest order $x_1x_2s=Q^2$ so
that the parton densities are probed at x values around $\sqrt{\tau}$. We have 
$\sqrt{\tau}=0.13-0.15$ at
$\sqrt{s}=630$ GeV (CERN
$Sp\bar pS$ Collider) and $\sqrt{\tau}=0.041-0.052$ at the Tevatron. In this respect the prediction is more delicate at the
LHC, where $\sqrt{\tau}\sim 5.9-6.5\cdot 10^{-3}$. One comparison of the experimental total rates at the Tevatron with the QCD
predictions is shown in Fig.~18, together with the expected rates at the LHC (based on the
structure functions obtained in \cite{rob}).

\begin{figure}[h]
\begin{center}
\includegraphics[width=8cm]{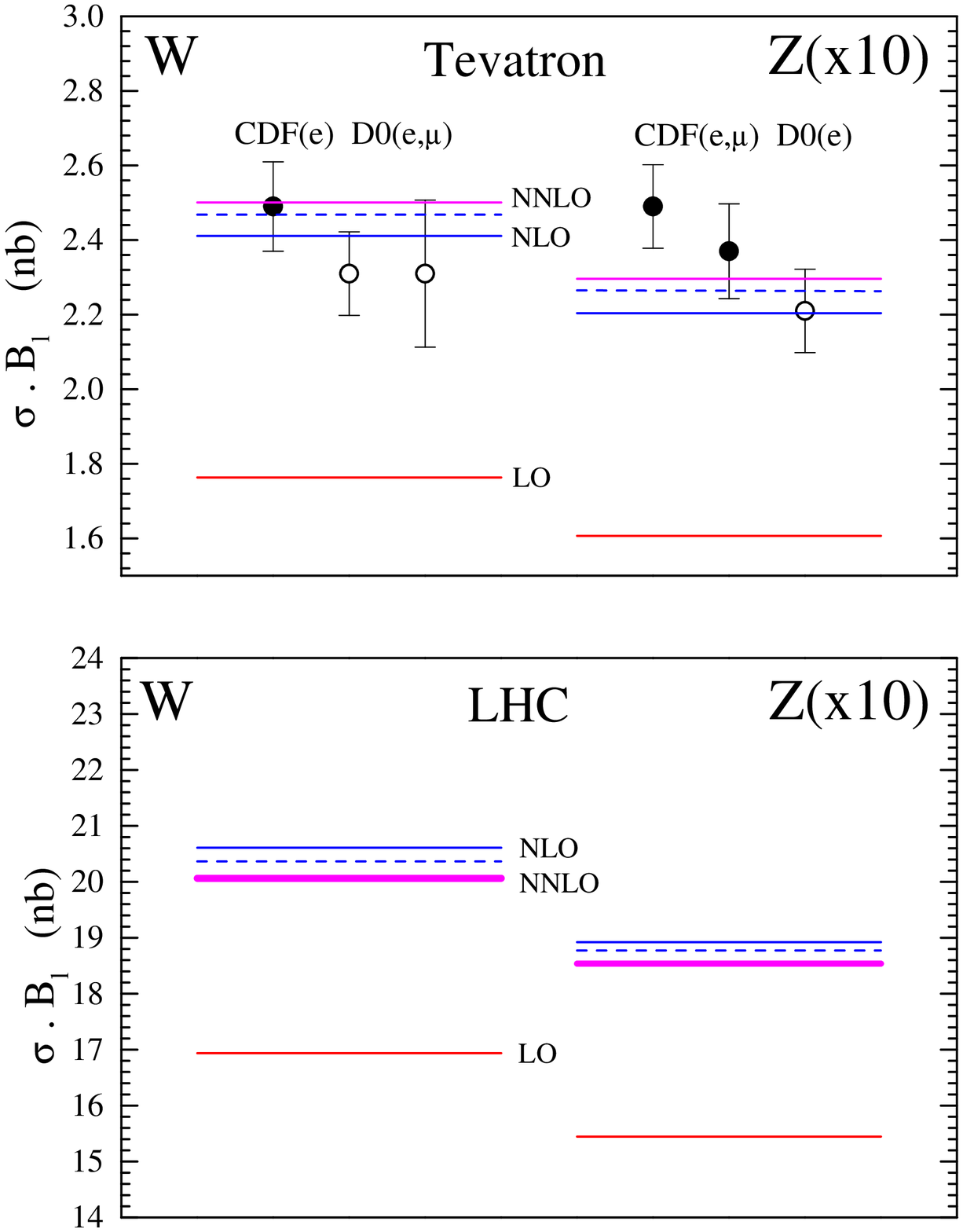}
\caption[]{Data vs. theory for $W$ and $Z$ production at the Tevatron ($\sqrt s$ = 1.8~TeV) together with the corresponding predictions for the LHC ($\sqrt s $= 1.4~TeV)}
\label{fig18}
\end{center}
\end{figure}  

The calculation of the W/Z $p_T$ distribution has been a classic problem in QCD. For large $p_T$, for example $p_T\sim
o(m_W)$, the $p_T$ distribution can be reliably computed in perturbation theory, which was done up to NLO in the late '70's
and early '80's. A problem arises in the intermediate range $\Lambda_{QCD}<<p_T<<m_W$, where the bulk of the data is
concentrated, because terms of order
$\alpha_s(p_T^2)\log{m_W^2/p_T^2}$ become of order 1 and should included to all orders. At order $\alpha_s$ we have:
\beq
\frac{1}{\sigma_0}\frac{d\sigma_0}{dp_T^2}~=~(1+A)\delta(p_T^2)~+~\frac{B}{p_T^2}\log{\frac{m_W^2}{p_T^2}}_+
~+~\frac{C}{(p_T^2)_+}~+~ D(p_T^2)\label{pT}\\
\eeq
where A, B, C, D are coefficients of order $\alpha_s$. The "+" distribution is defined in complete analogy with
eq.(\ref{plus}):
\beq
\int_0^{p^2_{TMAX}}g(z)f(z)_+dz~=~\int_0^{p^2_{TMAX}}[g(z)-g(0)]f(z)dz\label{pluspT}\\
\eeq
The content of this at first sight mysterious definition is that the singular "+" terms do not contribute to the total
cross-section. In fact for the cross-section the weight function $g(z)=1$ and we obtain:
\beq
\sigma~=~\sigma_0[(1+A)~+~\int_0^{p^2_{TMAX}}D(z)dz]\label{tot}\\
\eeq
The singular terms, of infrared origin, are present at the non completely inclusive level but disappear in the total
cross-section. These singularities are proven to exponentiate and lead to the following expression:
\beq
\frac{1}{\sigma_0}\frac{d\sigma_0}{dp_T^2}~=~\int \frac{d^2b}{4\pi}\exp{(-ib\cdot p_T)}(1+A)\exp{S(b)}\label{Sud}\\
\eeq
with:
\beq
S(b)~=~\int_0^{p_{TMAX}}\frac{d^2k_T}{2\pi}[\exp{ik_T\cdot b}-1][\frac{B}{k_T^2}\log{\frac{m_W^2}{k_T^2}}
~+~\frac{C}{k_T^2}]\label{s}\\
\eeq
At large $p_T$  the LO perturbative expansion is recovered. At intermediate $p_T$ the infrared $p_T$ singularities are
resummed (the Sudakov log terms, which are typical of vector gluons, are related to the fact that for a charged particle in
acceleration it is impossible not to radiate, so that the amplitude for no soft gluon emission is exponentially suppressed).
However this formula has problems at small $p_T$, for example, because of the presence of $\alpha_s$ under the integral for
$S(b)$: presumably the relevant scale is of order $k_T^2$. So it must be completed by some non perturbative ansatz or an
extrapolation into the soft region. All the exercise has been extended to NLO accuracy, where one starts from the
perturbative expansion at order
$\alpha_s^2$, and generalises the resummation to also include NLO terms of order $\alpha_s(p_T^2)^2\log{m_W^2/p_T^2}$. The
comparison with the data is very impressive. In Fig.~19 we see the $p_T$ distribution as predicted in QCD (with a number of
variants that mainly differ in the approach to the soft region) compared with some recent data from the D0 Collaboration at
the Tevatron \cite{D0}.

\begin{figure}[h]
\begin{center}
\includegraphics[width=8cm]{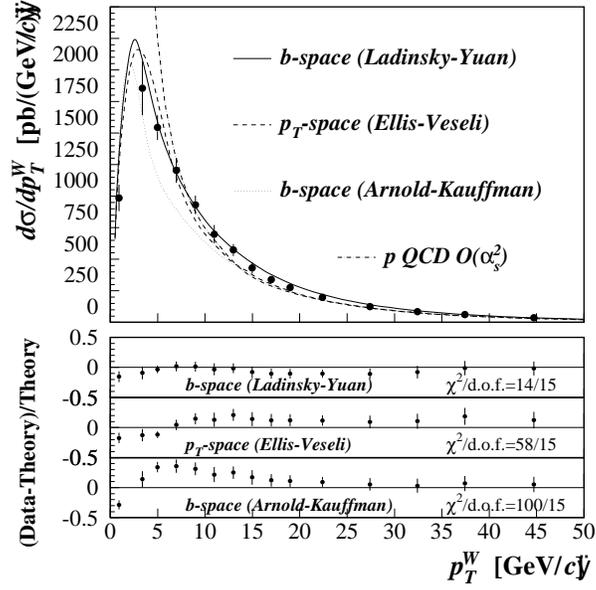}
\caption[]{QCD predictions for the $W~p_T$ distribution compared with recent D0 data at the Tevatron ($\sqrt s$ = 1.8 TeV)}
\label{fig19}
\end{center}
\end{figure}

\section{Measurements of $\alpha_s$}

The most precise and reliable measurements of $\alpha_s(m_Z^2)$ in the $\bar{MS}$ definition are obtained from $e^+e^-$
colliders (in particular LEP) and from Deep Inelastic Scattering. 

\subsection{$\alpha_s$ from $e^+e^-$ colliders}

The main methods at $e^+e^-$ colliders are: a) Inclusive hadronic Z decay, $R_l$, $\sigma_h$, $\Gamma_Z$. b) Inclusive
hadronic $\tau$ decay. c) Event shapes and jet rates.

As we have seen, for a quantity like $R_l$ we can write a general expression of the form:
\beq
R_l~=~\frac{Z,\tau\rightarrow hadrons)}{Z,\tau\rightarrow
leptons)}~\sim~R^{EW}(1~+~\delta_{QCD}~+~\delta_{NP})~+...\label{RR}\\
\eeq
where $R^{EW}$ is the electroweak-corrected Born approximation, $\delta_{QCD}$, $\delta_{NP}$ are the perturbative
(logarithmic) and non perturbative (power suppressed) QCD corrections. If we consider measurement at the Z, from $R_l$
only, assuming the standard electroweak theory, one finds \cite{lep}:
\beq
\alpha_s(m_Z)=0.123\pm0.004\label{alR}\\
\eeq
Better, one can use all info from $R_l$, $\Gamma_Z=3\Gamma_l+\Gamma_h$ and
$\sigma_h=12\pi\Gamma_l\Gamma_h/(m_Z^2\Gamma_Z^2)$. From these observables together one obtains: 
\beq
\alpha_s(m_Z)=0.120\pm0.003\label{alZ}\\
\eeq
By adding all other electroweak precision electroweak tests (in particular $m_W$) one finds:
\beq
\alpha_s(m_Z)=0.118\pm0.003\label{alEW}\\
\eeq
The final error is predominantly theoretical and is dominated by our ignorance on $m_H$ and from higher orders in the QCD
expansion.

We now consider the measurement of $\alpha_s(m_Z)$ from $\tau$ decay. $R_\tau$ has a number of advantages that, at least in
part, tend to compensate for the smallness of $m_\tau=1.777~GeV$. First, $R_\tau$ is maximally inclusive, more than
$R_{e^+e^-}(s)$, because one also integrates over all values of the invariant hadronic squared mass:
\beq
R_\tau=\frac{1}{\pi}\int_0^{m_\tau^2}\frac{ds}{m_\tau^2}(1-\frac{s}{m_\tau^2})^2 Im\Pi_\tau(s)\label{incl}\\
\eeq
Analyticity can be used to transform the integral into one on the circle at $|s|=m_\tau^2$:
\beq
R_\tau=\frac{1}{2\pi i}\oint_{|s|=m_\tau^2}\frac{ds}{m_\tau^2}(1-\frac{s}{m_\tau^2})^2 \Pi_\tau(s)\label{incl1}\\
\eeq  
Also, the factor $(1-\frac{s}{m_\tau^2})^2$ is important to kill the sensitivity the region $Res=m_\tau^2$ where the
physical cut and the associated thresholds are located. Still the quoted result (a combination of ALEPH and OPAL
analyses) looks a bit too precise:
\beq
\alpha_s(m_Z)=0.1181\pm0.0007(exp)\pm0.003(th)\label{altau}\\
\eeq
This precision is obtained by taking for granted that corrections suppressed by $1/m_\tau^2$ are negligible. This is
because, in the massless theory, the light cone expansion is given by:
\beq
\delta_{NP}=\frac{ZERO}{m_\tau^2}~+~c_4\cdot \frac{<O_4>}{m_\tau^4}~+~c_6\cdot \frac{<O_6>}{m_\tau^6}~+\cdots\label{OPEtau}\\
\eeq
In fact there are no dim-2 Lorentz and gauge invariant operators. For example, $g_{\mu}g^{\mu}$ is not gauge invariant. In
the massive theory, the ZERO is replaced by light quark mass-squared $m^2$. This is still negligible if $m$ is taken as a
lagrangian mass of a few MeV. But would not at all be negligible, actually would very much affect the result, if it is taken
as a constituent mass of order $m\sim\Lambda_{QCD}$. Most people believe the optimistic version. I am not convinced that
the gap is not filled up by ambiguities of $0(\Lambda_{QCD}^2/m_\tau^2)$ from $\delta_{pert}$. In any case, one can discuss
the error, but it is true and remarkable, that the central value from $\tau$ decay, obtained at very small $Q^2$, is in
perfect agreement with all other precise determinations of $\alpha_s$ at more typical LEP values of $Q^2$. 

\subsection{$\alpha_s$ from Deep Inelastic Scattering}

QCD predicts the $Q^2$ dependence of $F(x,Q^2)$ at each fixed x, not the x shape. But the $Q^2$ dependence is related to the
x shape by the QCD evolution equations. For each x-bin the data allow to extract the slope of an approximately stright line
in $dlogF(x,Q^2)/dlogQ^2$: the log slope. The $Q^2$ span and the precision of the data are not much sensitive to the
curvature, for most x values. A single value of $\Lambda_{QCD}$ must be fitted to reproduce the collection of the log
slopes. For the determination of $\alpha_s$ the scaling violations of non-singlet structure functions would be ideal,
because of the minimal impact of the choice of input parton densities. We can write the non-singlet evolution equations in
the form:
\beq
\frac{d}{dt}logF(x,t)~=~\frac{\alpha_s(t)}{2\pi}\int_x^1\frac{dy}{y}\frac{F(y,t)}{F(x,t)}P_{qq}(\frac{x}{y},\alpha_s(t))
\label{NSEE}\\
\eeq
where $P_{qq}$ is the NLO splitting function. It is clear from this form that, for example, the normalisation of the input
density drops away, and the dependence on the input is reduced to a minimum (also there is a single density, while in
general there are quark and gluon densities). Unfortunately the data on non-singlet structure functions are not very
accurate. If we take the difference of data on protons and neutrons, $F_p-F_n$, experimental errors add up in the difference
and finally are large. The $F_{3\nu N}$ data are directly non-singlet but are not very precise. A determination of
$\alpha_s$ from the CCFR data on $F_{3\nu N}$ has led to \cite{kat}:
\beq
\alpha_s(m_Z)=0.118\pm0.006\label{alCCFR}\\
\eeq
When one measures $\alpha_s$ from scaling violations on $F_2$ from e or $\mu$ beams, the data are abundant, the errors
small but there is an increased dependence on input parton densities and especially a strong correlation between the result
on $\alpha_s$ and the input on the gluon density. There are two most complete and accurate derivations of $\alpha_s$ from
scaling violations in $F_2$. In the first analysis, by Santiago and Yndurain \cite{ynd}, the data on protons from SLAC,
BCDMS, E665 and HERA are used with NLO kernels plus the NNLO first few moments. The analysis is based on an original method
that uses projections on a specially selected basis of orthogonal polynomials. The quoted result is given by:
\beq
\alpha_s(m_Z)=0.117\pm0.003\label{firstalSY}\\
\eeq
A different analysis by Alekhin \cite{ale} of a similar collection of proton data from SLAC, BCDMS, NMC and
HERA with NLO kernels and a more conventional method leads to
\beq
\alpha_s(m_Z)=0.116\pm0.003\label{alSY}\\
\eeq
In both analysis the dominant error is theoretical and could be somewhat larger than quoted. 

If we compare these results on
$\alpha_s$ from DIS with the findings at the Z, given by eq.(\ref{alEW}), we see that the agreement looks perfect. But, in
my opinion, the situation of $\alpha_s$ from DIS is not yet completely satisfactory (while it is so for $\alpha_s$ from
$e^+e^-$). The data have shown large fluctuations in the recent past. For example, the original result obtained by
combining BCDMS and SLAC data was \cite{vir}: 
\beq
\alpha_s(m_Z)=0.113\pm0.005\label{alBCDMS}\\
\eeq
This low value appeared to be confirmed by the CCFR result from $F_2$ and $F_3$ data combined:
\beq
\alpha_s(m_Z)=0.111\pm0.005\label{alCCFRold}\\
\eeq
But later the same data were corrected by a new energy calibration performed by the collaboration and the result become:
\beq
\alpha_s(m_Z)=0.119\pm0.005\label{alCCFRnew}\\
\eeq
So the experimental errors are perhaps larger than quoted and there is a problem for matching the systematics of different
experiments. On the theory side the analysis methods are perhaps still not completely optimised.

Summarising: there is very good agreement among many different measurements of $\alpha_s$ (see Fig.~20, \cite{pdg}). This is a
very convincing, quantitative test of QCD. The average value quoted by PDG 2000 is 
\beq
\alpha_s(m_Z)=0.118\pm0.002\label{alPDG}\\
\eeq

\begin{figure}[h]
\begin{center}
\includegraphics[width=7cm]{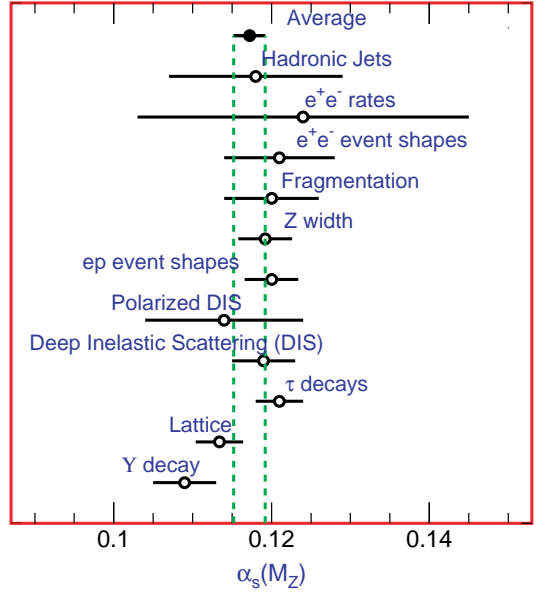}
\caption[]{The PDG (Ref.~\cite{pdg}) compilation of the existing precise measurements of $\alpha_s(M_2)$}
\label{fig20}
\end{center}
\end{figure} 

The corresponding value of $\Lambda$ (for $n_f=5$) is:
\beq
\Lambda_5=209\pm24~MeV\label{lambda}\\
\eeq
$\Lambda$ is the scale of mass that finally appears in massless QCD. It is the scale where $\alpha_s(\Lambda)$ is of order
1. Hadron masses are determined by $\Lambda$. Actually the $\rho$ mass or the nucleon mass receive little contribution from
the quark masses (the case of pseudoscalar mesons is special, as they are the pseudo Goldstone bosons of broken chiral
invariance). Hadron masses would be almost the same in massless QCD. 

\section{Conclusion}

We have seen that perturbative QCD based on asymptotic freedom offers a rich variety of tests and we have described some
examples in detail. QCD tests are not as precise as for the electroweak sector. But the number and diversity of such tests
has established a very firm experimental foundation for QCD as a theory of strong interactions.

The field is still very much in movement. There are areas of continuous development: I list here some of these domains
where work is in progress and which would deserve an expanded discussion. 

¥ Higher order calculations: important examples are the effort to complete the 3-loop analytic determination of the
splitting functions, and the calculation of 4-jet distributions at NLO in $e^+e^-$ annihilation.

¥ Resummations in problems with 2 (or more) different large scales: if y=scale1/scale2 one needs to  resum
$(\alpha_s\log^a{y})^n$, with
$a=1,2$, at all orders in n. We have seen an example on the $W,Z$ $p_T$ distribution in $pp$ or $p\bar p$ collisions for 
$\Lambda_{QCD}<<p_T<<m_W$ where all terms of order $(\alpha_s(p_T^2)\log{m_W^2/p_T^2})^n$ must be taken into account. Other
examples involve structure functions at small x ($(\alpha_s(Q^2)\log{1/x})^n$) or at $x\sim1$
($(\alpha_s(Q^2)\log{1/(1-x)})^n$), or thrust distributions near $T\sim 1$ etc.

¥ Renormalons and power suppressed corrections (for an introduction see, for example, Ref~\cite{sixteen}).. The QED and QCD perturbative series, after renormalisation, have all their
coefficients finite, but it is well known that the expansion does not converge. Actually the perturbative series is not
even Borel summable. After Borel resummation for a given process one is left with a result which is ambiguous by terms
typically down by $-\exp{(-n/b\alpha)}$, with n an integer and b the first $\beta$ function coefficient. In QED these
corrective terms are extremely small and not very important in practice. On the contrary in QCD $\alpha=\alpha_s(Q^2)\sim
1/(b\log{Q^2/\Lambda^2})$ and the ambiguous terms are of order $(1/Q^2)^n$, that is are power suppressed. The problem
arises of the precise relation between the ambiguities of the perturbative expansion and the higher twist corrections, which
is a very interesting field of research.

¥ New areas explored by experiment. Particularly interesting physics cases are present in polarized deep inelastic
scattering, or in the phenomenology of structure functions at small x as measured at HERA.

¥ An important domain of activity is the QCD event simulation for the preparation of LHC experiment which poses highly non
trivial theoretical and technical problems.

\section{APPENDIX: The Formalism of Gauge Theories}

We summarize here the definition and the structure of a gauge Yang--Mills theory. 

Consider a Lagrangian density ${\cal L}[\phi,\partial_{\mu}\phi]$ which is invariant under a $D$ dimensional continuous
group of transformations:
\begin{equation}
\phi' = U(\theta^A)\phi\quad\quad (A = 1, 2, ..., D)~.
\label{7}
\end{equation} For $\theta^A$ infinitesimal, $U(\theta^A) = 1 + ig \sum_A~\theta^AT^A$, where
$T^A$ are the generators of the group $\Gamma$ of transformations (\ref{7}) in the (in general reducible)
representation of the fields $\phi$. Here we restrict ourselves to the case of internal symmetries, so that $T^A$ are
matrices that are independent of the space--time coordinates. The generators $T^A$ are normalized in such a way that
for the lowest dimensional non-trivial representation of the group $\Gamma$ (we use $t^A$ to denote the generators in
this particular representation) we have
\begin{equation} {\rm tr}(t^At^B) = \frac{1}{2} \delta^{AB}~.
\label{8}
\end{equation} The generators satisfy the commutation relations
\begin{equation} [T^A,T^B] = iC_{ABC}T^C~.
\label{9}
\end{equation} In the following, for each quantity $V^A$ we define
\begin{equation} {\bf V} = \sum_A~T^AV^A~.
\label{10}
\end{equation} If we now make the parameters $\theta^A$ depend on the space--time coordinates
$\theta^A = \theta^A(x_{\mu}),$ ${\cal L}[\phi,\partial_{\mu}\phi]$ is in general no longer invariant under the gauge
transformations $U[\theta^A(x_{\mu})]$, because of the derivative terms. Gauge invariance is recovered if the ordinary
derivative is replaced by the covariant derivative:
\begin{equation} D_{\mu} = \partial_{\mu} + ig{\bf V}_{\mu}~,
\label{11}
\end{equation} where $V^A_{\mu}$ are a set of $D$ gauge fields (in one-to-one correspondence with the group generators)
with the transformation law
\begin{equation} {\bf V}'_{\mu} = U{\bf V}_{\mu}U^{-1} - (1/ig)(\partial_{\mu}U)U^{-1}~.
\label{12}
\end{equation} For constant $\theta^A$, {\bf V} reduces to a tensor of the adjoint (or regular) representation of the
group:
\begin{equation} {\bf V}'_{\mu} = U{\bf V}_{\mu}U^{-1} \simeq {\bf V}_{\mu} + ig[\theta, {\bf V}_{\mu}]~,
\label{13}
\end{equation} which implies that
\begin{equation} V'^C_{\mu} = V^C_{\mu} - gC_{ABC}\theta^AV^B_{\mu}~,
\label{14}
\end{equation} where repeated indices are summed up.

As a consequence of Eqs. (\ref{11}) and (\ref{12}), $D_{\mu}\phi$  has the same transformation pro\-perties as $\phi$:
\begin{equation} (D_{\mu}\phi)' = U(D_{\mu}\phi)~.
\label{15}
\end{equation}

Thus ${\cal L}[\phi,D_{\mu}\phi]$ is indeed invariant under gauge transformations. In order to construct a
gauge-invariant kinetic energy term for the gauge fields $V^A$, we consider
\begin{equation} [D_{\mu},D_{\nu}] \phi =  ig\{\partial_{\mu}{\bf V}_{\nu} - \partial_{\nu}{\bf V}_{\mu} + ig[{\bf
V}_{\mu},{\bf V}_{\nu}]\}\phi \equiv ig {\bf F}_{\mu\nu}\phi~,
\label{16}
\end{equation} which is equivalent to
\begin{equation} F^A_{\mu\nu} = \partial_{\mu}V^A_{\nu} - \partial_{\nu}V^A_{\mu} - gC_{ABC}V^B_{\mu}V^C_{\nu}~.
\label{17}
\end{equation} From Eqs. (\ref{7}), (\ref{15}) and (\ref{16}) it follows that the transformation properties of
$F^A_{\mu\nu}$ are those of a tensor of the adjoint representation
\begin{equation} {\bf F}'_{\mu\nu} = U{\bf F}_{\mu\nu}U^{-1}~.
\label{18}
\end{equation} The complete Yang--Mills Lagrangian, which is invariant under gauge transformations, can be written in
the form
\begin{equation} {\cal L}_{\rm YM} = - \frac{1}{4} \sum_A F^A_{\mu\nu}F^{A\mu\nu} + {\cal L} [\phi,D_{\mu}\phi]~.
\label{19}
\end{equation}

For an Abelian theory, as for example QED, the gauge transformation reduces to
$U[\theta(x)] = {\rm exp} [ieQ\theta(x)]$, where $Q$ is the charge generator. The associated gauge field (the photon),
according to Eq. (\ref{12}), transforms as
\begin{equation} V'_{\mu} = V_{\mu} - \partial_{\mu}\theta(x)~.
\label{20}
\end{equation} In this case, the $F_{\mu\nu}$ tensor is linear in the gauge field $V_{\mu}$ so that in the absence of
matter fields the theory is free. On the other hand, in the non-Abelian case the $F^A_{\mu\nu}$ tensor contains both
linear and quadratic terms in $V^A_{\mu}$, so that the theory is non-trivial even in the absence of matter fields.

\end{document}